\documentclass[12pt]{article}
\usepackage{graphicx}
\usepackage{epstopdf}

\begin{document}
\baselineskip 0.5825cm
\newcommand{\tri}{\triangleright}
\newcommand{\range}{{\rm range}}
\newcommand{\Ree}{{\rm Re }}
\newcommand{\Imm}{{\rm Im }}
\newcommand{\diag}{{\rm diag}}
\newcommand{\sign}{{\rm sign}}
\newcommand{\tr}{{\rm tr}}
\newcommand{\rank}{{\rm rank}}
\newcommand{\bp}{\bigskip}
\newcommand{\mdp}{\medskip}
\newcommand{\slp}{\smallskip}
\newcommand{\Rw}{\Rightarrow}
\newcommand{\ts}{& \hspace{-0.1in}}
\newcommand{\nn}{\nonumber}
\newcommand{\bea}{\begin{eqnarray}}
\newcommand{\eea}{\end{eqnarray}}
\newcommand{\beas}{\begin{eqnarray*}}
\newcommand{\eeas}{\end{eqnarray*}}
\newcommand{\beq}{\begin{equation}}
\newcommand{\eeq}{\end{equation}}
\newtheorem{exa}{Example}[section]
\newtheorem{thm}{Theorem}[section]
\newtheorem{lem}{Lemma}[section]
\newtheorem{prop}{Proposition}[section]
\newtheorem{fact}{Fact}[section]
\newtheorem{cor}{Corollary}[section]
\newtheorem{defn}{Definition}[section]
\newtheorem{rem}{Remark}[section]
\renewcommand{\theequation}{\thesection.\arabic{equation}}

\title{{\bf Discrete-Time Adaptive State Tracking Control Schemes
    Using Gradient Algorithms}}
\author{{\it Gang Tao}\\
\normalsize Department  of Electrical and Computer Engineering\\  
\normalsize University of Virginia \\ 
\normalsize Charlottesville, VA 22903, USA} 
\date{} 
\maketitle 
\begin{abstract}
This paper conducts a comprehensive study of a classical adaptive
control problem: adaptive control of a state-space plant model:
$\dot{x}(t) = A x(t) + B u(t)$ in continuous time, or 
$x(t+1) = A x(t) + B u(t)$ in discrete time, for state tracking of a
chosen stable reference model system: $\dot{x}_m(t) = A_m x_m(t)
+ B_m r(t)$  in continuous time, or $x_m(t+1) = A_m x_m(t) + B_m
r(t)$ in discrete time. 
Adaptive state tracking control schemes for continuous-time systems
have been reported in the literature, using a Lyapunov design and
analysis method which has not been successfully applied to
discrete-time systems, so that the discrete-time adaptive state
tracking problem has remained to be open. In this paper, new adaptive
state tracking control schemes are developed for discrete-time
systems, using a gradient method for the design of adaptive laws for
updating the controller parameters. Both direct and indirect adaptive
designs are presented, which have the standard and desired adaptive
law properties. Such a new gradient algorithm based framework is also
developed for adaptive state tracking control of continuous-time
systems, as compared with the Lyapunov method based framework.
\end{abstract}



\section{Introduction}
Adaptive control is a methodology for feedback control of dynamic
systems with parameter, structure, actuator and sensor uncertainties. 
Parametrized system structure, actuator and sensor uncertainties can
be dealt with by adaptive control schemes effectively. Of the broad
areas of adaptive control research (see for example,
\cite{af21}-\cite{zsc23}, adaptive state tracking control is one topic
of special interests. It meets certain desired system performance that
the controlled plant state vector asymptotically tracks the reference
model system state vector. It has a complete, direct and straightforward
Lyapunov design and analysis framework for the continuous-time
adaptive control case, under a necessary plant-model matching condition
\cite{is96}, \cite{na89}, \cite{t03}, \cite{t14}. 

Such a Lyapunov method has two important features: the first feature is
that a positive definite function $V$ containing both the system state
tracking error $e(t)$ and the parameter error $\tilde{\theta}(t)$ is used for
system stability analysis. The second feature is that the adaptive
laws are chosen to make the time-derivative $\dot{V} \leq - e^T(t) e(t)
\leq 0$, from which the system stability is ensured in the Lyapunov
sense (making $V$ a Lyapunov function of the adaptive control system)
and the state tracking error $e(t)$ is ensured, directly via Barbalat
lemma, to be $\lim_{t \rightarrow \infty} e(t) = 0$. Such a Lyapunov
method based straightforward design and analysis framework has in the
recent years attracted researchers to pursue its extensions and
applications for new adaptive control developments 
(see, for example, \cite{lw13} and the literature review in \cite{af21}). 

\medskip
However, the discrete-time adaptive state tracking control
problem has remained to be open, as the Lyapunov method based
framework has not been successfully applied to adaptive state tracking
control of discrete-time systems with unknown parameters. In fact,
there is no visible study in the literature on such a problem.

In this paper, we develop a new gradient algorithm based framework and its
derived adaptive control schemes, to solve this long-standing discrete-time
adaptive state tracking control problem, and to provide new solutions
to continuous-time adaptive state tracking control. New features of
such a framework include the formation of common-parameter estimation
errors and development of composite-error adaptive laws. We conduct a
comprehensive study of such classical and yet open adaptive control
problems, derive adaptive control schemes of different types and
for different systems, and show their stability analysis and
comparisons. 

\medskip
In Section 2, we present the adaptive state tracking control problems
and review the direct and indirect adaptive control schemes for
continuous-time systems, based on a Lyapunov method. In Section 3, we
develop a gradient algorithm based framework for solving the
discrete-time adaptive state tracking control problem, with both the
direct and indirect adaptive control schemes. In Section 4, we develop
the gradient algorithm based framework for adaptive state tracking control
of systems with multiple inputs, and in particular, we derive direct
and indirect adaptive control schemes for solving the open
discrete-time problems. In Sections 3 and 4, we also demonstrate how
such a gradient algorithm based framework can be used for
continuous-time systems.
 
\setcounter{equation}{0}
\section{Adaptive State Tracking Control}
In this section, we first formulate the adaptive state tracking
control problems, and then give an overview of adaptive state tracking
control designs for continuous-time systems.

\subsection{Problem Formulation}
Consider the continuous-time single-input multi-output (SIMO) plant
\beq
\dot{x}(t) = A x(t) + b u(t),\;
x(t) \in R^{n},\;u(t) \in R,
\label{4plant}
\eeq
where $A \in R^{n \times n}$, $b \in R^{n}$ are unknown constant
parameters, and assume that the state (output) vector $x(t)$ is
available for measurement. The state tracking control objective is to
design a feedback control law for the plant input $u(t)$ such that all
 closed-loop system signals are bounded and
$x(t)$ asymptotically tracks a reference state vector
$x_{m}(t)$ of a chosen reference model system
\beq
\dot{x}_{m}(t) = A_{m} x_{m}(t) + b_{m} r(t),\;
x_{m}(t) \in R^{n},\;r(t) \in R,
\label{4rms}
\eeq
where $A_{m} \in R^{n \times n}$, $b_{m} \in R^{n}$
with $A_m$ stable in the continuous-time sense: all eigenvalues of
$A_m$ are the open left-half of the complex $s$-plane (for desired reference model
system stability and performance), and $r(t)$ is a chosen bounded
reference input signal for desired system response. 

\medskip
For the discrete-time counterparts, the plant model is
\beq
x(t+1) = A x(t) + b u(t),\;
x(t) \in R^{n},\;t = 0, 1, 2, \ldots, 
\label{432A.1}
\eeq
where $A \in R^{n \times n}$ and $b \in R^{n}$ are 
unknown constant matrix and vector, $x(t) \in R^n$ is the
plant state (output) vector, and $u(t) \in R$ is the input signal, and
the reference model system is
\beq
x_{m}(t+1) = A_{m} x_{m}(t) + b_{m} r(t),\;
x_{m}(t) \in R^{n},\;r(t) \in R
\label{432A.2}
\eeq
where $A_{m} \in R^{n \times n}$ and $b_{m} \in R^{n}$ are some 
constant matrix and vector, all eigenvalues of $A_{m}$ are
inside the unit circle of the complex $z$-plane, and $r(t)$ is a chosen 
bounded reference input signal.

\medskip
The following basic assumptions are needed to solve such a state
tracking problem.

\begin{description}
\item[] {\bf Assumption (A1)}: There exist a constant vector $k_{1}^{*} \in R^{n}$ and a nonzero
 constant scalar $k_{2}^{*} \in R$ such that the following equations are
satisfied:
\beq
A + b k_{1}^{*T} = A_{m},\;
b k_{2}^{*} = b_{m}.
\label{4me}
\eeq

\item[] {\bf Assumption (A2)}: The sign of $k_{2}^{*}$, $\sign[k_{2}^{*}]$, is known.
\end{description}

Such an assumption (\ref{4me}) is needed not only for solving the adaptive control problem
when the plant parameters $A$ and $b$ are unknown, but also for solving the
nominal control problem when $A$ and $b$ are known.
 
\subsection{Continuous-Time Adaptive Control Designs}
\label{4Continuous-Time Designs}
The state feedback state tracking controller structure is
\beq
u(t) = k^{T}_{1}(t) x(t) + k_{2}(t) r(t),
\label{432A.7}
\eeq
where $k_{1}(t)$ and $k_{2}(t)$ are the estimates of $k_{1}^{*}$ and
$k_{2}^{*}$ satisfying Assumption (A1). There are two methods to
design an adaptive scheme to update the controller parameters: a
direct method to adaptively update $k_1(t)$ and $k_2(t)$ directly, and
an indirect method to adaptively update the estimates of the plant
parameters and then to calculate the controller parameters $k_1(t)$
and $k_2(t)$ from the plant parameter estimates.

\subsubsection{Direct Adaptive Control Design}
With the control law (\ref{432A.7}), the plant (\ref{4plant}) becomes
\bea
\dot{x}(t) \ts = \ts A x(t) + b (k^{T}_{1}(t) x(t) + k_{2}(t) r(t)) \nn\\
\ts = \ts A x(t) + b (k_1^{*T} x(t) + k_{2}^* r(t)) + b (\tilde{k}_1^{T}(t) x(t) + \tilde{k}_{2}(t) r(t)) \nn\\
\ts = \ts A_m x(t) + b_m r(t) + b (\tilde{k}_1^{T}(t) x(t) + \tilde{k}_{2}(t) r(t)),
\label{4plantcc}
\eea
where $A_m = A + b k_{1}^{*T}$ and $b_m = b k_{2}^{*}$ (see Assumption
(A1)), and 
\beq
\tilde{k}_1(t) = k_1(t) - k_1^*,\;\tilde{k}_{2}(t) = k_2(t) - k_2^*.
\eeq
With (\ref{4rms}) and $e(t) = x(t) - x_m(t)$, we have
\beq
\dot{e}(t) = A_m e(t) + b_m \frac{1}{k_2^*} (\tilde{k}_1^{T}(t) x(t)
+ \tilde{k}_{2}(t) r(t)).
\label{4edot}
\eeq

\medskip
{\bf Adaptive laws}. We choose the adaptive laws for $k_{1}(t)$ and $k_{2}(t)$ as
\bea
\label{4k1dotc}
\dot{k}_{1}(t) \ts = \ts - \sign[k_{2}^{*}] \Gamma x(t) e^{T}(t) P b_{m}\\*[0.05in]
\dot{k}_{2}(t) \ts = \ts - \sign[k_{2}^{*}] \gamma r(t) e^{T}(t) P b_{m},
\label{4k2dotc}
\eea
where $\Gamma = \Gamma^T > 0$ and $\gamma > 0$ are chosen adaptation
gains, and $P = P^{T} > 0$ satisfying $P A_{m} + A_{m}^{T} P = - Q$,
for a chosen constant $n \times n$ matrix $Q = Q^{T} > 0$.

\medskip
Consider the positive definite function for this Lyapunov-type algorithm: 
\beq
V = e^{T} P e + \frac{1}{|k_{2}^{*}|} \tilde{k}_{1}^{T} \Gamma^{-1}
\tilde{k}_{1} + \frac{1}{|k_{2}^{*}|} \tilde{k}_{2}^{2} \gamma^{-1}
\label{432A.400}
\eeq
and its time-derivative 
\beq
\dot{V} = - e^{T}(t) Q e(t) \leq 0,
\label{4Vdotc}
\eeq
we conclude that $\tilde{k}_1(t) = k_1(t) - k_1^*$, $\tilde{k}_2(t) =
k_2(t) - k_2^*$ and $e(t) = x(t) - x_m(t)$ are all bounded, and $e(t)
\in L^2$. We further have from (\ref{432A.7}) that $u(t)$ is bounded
and from (\ref{4edot}) that $\dot{e}(t)$ is bounded so that $\lim_{t
  \rightarrow \infty} e(t) = 0$ (following Barbalat lemma
\cite{is96}). This result is summarized as:

\begin{prop}
\label{4prop}
The control law (\ref{432A.7}), updated by the adaptive laws
(\ref{4k1dotc})-(\ref{4k2dotc}) and applied to the plant
(\ref{4plant}), ensures the closed-loop system signal boundedness and
asymptotic tracking: $\lim_{t \rightarrow \infty} (x(t) - x_m(t)) = 0$.
\end{prop}

\subsubsection{Indirect Adaptive Control Design}
\label{4Indirect Adaptive Designc}
Using Assumption (A1), we parametrize the plant (\ref{4plant}) as
\bea
\dot{x}(t) \ts = \ts A x(t) + b u(t) = A_m x(t) + b_m (\theta_2^{*} u(t) -
\theta_1^{*T} x(t)),
\label{4para}
\eea
where $\theta_1^* = k_2^{*-1} k_1^*$ and $\theta_2^* = k_2^{*-1}$, 
and, with the estimates $\theta_1(t)$ and $\theta_2(t)$ of
$\theta_1^*$ and $\theta_2^*$, 
generate an {\it a posteriori} estimate $\hat{x}(t)$ of $x(t)$
from the estimator equation 
\beq \dot{\hat{x}}(t) = A_m \hat{x}(t) + b_m
(\theta_2(t) u(t) - \theta_1^{T}(t) x(t)). \label{4est} \eeq 
For $e_x(t) = \hat{x}(t) - x(t)$, we have the
estimator state error equation \beq \dot{e}_x(t) = A_m e_x(t) + b_m
((\theta_2(t) -
\theta_2^*) u(t) - (\theta_1(t) - \theta_1^*)^T x(t)). \label{4eee}
\eeq

\medskip
{\bf Adaptive laws}. 
Then, we choose the adaptive laws for $\theta_1(t)$ and $\theta_2(t)$:
\bea
\dot{\theta}_1(t) \ts  = \ts \Gamma_{1} x(t) e_x^T(t)P b_m \label{4theta1d}\\
\dot{\theta}_2(t) \ts = \ts - \gamma_{2} u(t) e_x^T(t) P b_m, \label{4theta2d}
\eea
where $\Gamma_{1} = \Gamma_{1}^{T} > 0$, $\gamma_{2} > 0$, and
$P = P^{T} > 0$ such that $P A_{m} + A_{m}^{T} P = - Q$ for a chosen
$Q = Q^{T} > 0$. 

\medskip
For the positive definite function
\beq
V = e_x^T P e_x + (\theta_1 - \theta_1^*)^T\Gamma_1^{-1}
(\theta_1 - \theta_1^*) + (\theta_2 - \theta_2^*)^2
\gamma_2^{-1},
\label{432A.40ex}
\eeq
we derive its time-derivative as: 
$
\dot{V} = - e_x^T Q e_x \leq 0,
$
so that $\theta_1(t)$, $\theta_2(t)$ and $e_x(t)$ are
all bounded, and $e_x(t) \in L^2$, as the basic properties of the
adaptive laws (\ref{4theta1d})-(\ref{4theta2d}).
\footnote{A vector signal $x(t)$ belongs to $L^\infty$: $x(t) \in
L^\infty$, if $x(t)$ is bounded, and $x(t) \in L^2$ if 
$\int_{0}^\infty x^T(t) x(t) \,dt < \infty$ (in the continuous-time
case) or $\sum_{t=0}^\infty x^T(t) x(t) < \infty$ (in the
discrete-time case).}

\medskip
\medskip
{\bf Control law}. We choose the adaptive control law
\beq
u(t) = \frac{1}{\theta_2(t)} v(t),\;v(t) = \theta_1^T(t) x(t) + r(t).
\label{4acl}
\eeq

To implement this control law, $\theta_2(t)$
should be projected to be away from zero:
$|\theta_2(t)| \geq \theta_2^a > 0$. This can be done using some
information about $k_2^*$, described by:

\begin{description}
\item[] {\bf Assumption (A3)}: An upper bound $k_2^b$ of
$|k_2^*|$: $k_2^b \geq |k_2^*|$, is known.
\end{description}

Then, $\theta_2^a = 1/k_2^b$ is a lower bound of
$|\theta_2^*|$: $0 < \theta_2^a \leq |\theta_2^*|$, as $\theta_2^* =
1/k_2^*$. With Assumption (A3), the adaptive law (\ref{4theta2d}) is modified as
\beq
\dot{\theta}_2(t) =  - \gamma_{2} u(t) e_x^T(t) P b_m +
f_2(t),\;\sign[\theta_2^*] \theta_2(0) \geq \theta_2^a,
\eeq
where $f_2(t)$ is a parameter projection signal: for $g_2(t) =
- \gamma_{2} u(t) e_x^T(t) P b_m$, 
\beq
f_{2}(t) = \left\{ \begin{array}{ll}
0  & \mbox{if $\sign[\theta_2^*]\theta_{2}(t) > \theta_2^a$, or}\\
 & \mbox{if $\sign[\theta_2^*] \theta_{2}(t) = \theta_2^a$ and
 $\sign[\theta_2^*] g_{2}(t) \geq 0$}\\*[0.05in]
- g_{2}(t) & \mbox{otherwise,}
\end{array}
\right.
\label{4f2(t)}
\eeq
which ensures that $\sign[\theta_2(t)] =
\sign[\theta_2^*]$, $|\theta_2(t)| \geq \theta_2^a > 0$ and 
$(\theta_2(t) - \theta_2^*) f_2(t) \leq 0$.

\medskip
With the control law (\ref{4acl}),
the estimator state equation (\ref{4est}) becomes
\beq
\dot{\hat{x}}(t) = A_m \hat{x}(t) + b_m r(t)
\label{4est1}
\eeq
which implies that $\hat{x}(t)$ is bounded so that $x(t)$ and
$\dot{x}(t)$ are bounded as $e_x(t) = \hat{x}(t) - x(t)$ is bounded,
and that $\lim_{t \rightarrow \infty} (\hat{x}(t) - x_m(t)) = 0$
exponentially so that $\hat{x}(t) - x_m(t) \in L^2$. Hence we have $x(t) -
x_m(t) = \hat{x}(t) - x_m(t) + x(t) - \hat{x}(t) \in L^2$ as $x(t) -
\hat{x}(t) \in L^2$, and finally, following Barbalat lemma, we have that
 $\lim_{t \rightarrow \infty} (x(t) - x_m(t)) = 0$. In summary, we have:

\begin{prop}
\label{4Prop1}
The adaptive controller (\ref{4acl}), updated from the adaptive law
(\ref{4theta1d})--(\ref{4theta2d}) and applied to the plant
(\ref{4plant}), ensures that all closed-loop system signals are bounded
and $\lim_{t \rightarrow \infty} (x(t) - x_m(t)) = 0$.
\end{prop}

The above direct and indirect adaptive control schemes are based on a
Lyapunov method characterized by two features. The first feature is that 
the positive definite function $V$ in (\ref{432A.400}) or
(\ref{432A.40ex}) contains the state error signal $e(t) = x(t) -
x_m(t)$ or $e_x(t) = \hat{x}(t) - x(t)$, in addition to the parameter
errors $\tilde{k}_1(t) = k_1(t) - k_1^*$ and $\tilde{k}_2(t) = k_2(t)
- k_2^*$ or $\tilde{\theta}_1(t) = \theta_1(t) - \theta_1^*$ and
$\tilde{\theta}_2(t) = \theta_2(t) - \theta_2^*$. The second feature
is that the adaptive laws are chosen to ensure the desired
time-derivative of $V$: $\dot{V} = - e^T Q e \leq 0$ or $\dot{V} = -
e_x^T Q e_x \leq 0$, from which the adaptive system stability directly
follows, making the adaptive control system stable in the Lyapunov
stability sense and $e(t) \in L^2$ or $e_x(t) \in L^2$, explicitly leading to 
 $\lim_{t \rightarrow \infty} e(t) = 0$ or 
 $\lim_{t \rightarrow \infty} e_x(t) = 0$.

Such an adaptive control design and analysis framework may be called a Lyapunov method
based framework, as the positive definite function $V$ is a Lyapunov
function of the adaptive control system: it contains the full error
signals and it is ensured to be $\dot{V} \leq 0$. It however has not
been successfully applied for adaptive state tracking control of
discrete-time systems, more precisely, it has not
been verified that a certain choice of adaptive laws can make 
such a positive definite function $V$ a nonincreasing function for a
discrete-time adaptive control system.
 
\setcounter{equation}{0}
\section{Discrete-Time Adaptive Control Designs}
\label{4Discrete-Time Adaptive Control Designs}
In this section, we develop two new adaptive control schemes: one
direct design and one indirect design, using gradient algorithms, to
solve the long-standing open discrete-time adaptive state tacking
control problem, with system stability and tracking performance
analysis and illustration. 

\medskip
Consider the single-input multi-output (SIMO) time-invariant plant
(\ref{432A.1}): $x(t+1) = A x(t) + b u(t)$, and the reference model
system (\ref{432A.2}): $x_{m}(t+1) = A_{m} x_{m}(t) + b_{m} r(t)$,
satisfying the conditions of Assumption (A1): $A + b k_{1}^{*T} = A_{m},\;
b k_{2}^{*} = b_{m}$, for some constant $k_{1}^{*} \in R^{n}$ and $k_{2}^{*} \in R$.

\medskip
\medskip
{\bf Nominal control}. With the parameters $k_1^*$ and $k_2^*$
satisfying Assumption (A1), the control law
\beq
u(t) = k^{*T}_{1} x(t) + k_{2}^{*} r(t)
\label{432A.4}
\eeq
can achieve the desired control objective: the
closed-loop control system with (\ref{432A.4}) becomes
\beq
x(t+1) = A x(t) + b (k^{*T}_{1} x(t) + k_{2}^{*} r(t)) = 
A_{m} x(t) + b_{m} r(t)
\label{432A.5}
\eeq
so that the plant state vector $x(t)$ is bounded, and so is the
control $u(t)$ in (\ref{432A.4}), and the tracking error 
$e(t) = x(t) - x_{m}(t)$ satisfies
\beq
e(t+1) = A_{m} e(t),\;e(0) = x(0) - x_{m}(0)
\label{432A.6}
\eeq
leading to $\lim_{t \rightarrow \infty} e(t) = 0$ exponentially. Such
a control law is called the nominal control law.

It is clear that the condition (\ref{4me}) is also necessary for the
control law (\ref{432A.4}) to achieve the control objective, even if the
parameters $A$ and $b$ are known. The condition  (\ref{4me}) is
the so-called matching condition for the closed-loop control system to
match the reference model system (\ref{432A.2}), exponentially with a
nominal controller for the known plant parameter case, as shown above,
or asymptotically with an adaptive controller for the unknown plant
parameter case, as shown next.

\begin{rem}
\rm
We note that the Lyapunov method used for the continuous-time
adaptive state feedback state tracking system design and analysis in
Section \ref{4Continuous-Time Designs} has not been shown to be
applicable to its discrete-time counterpart whose problem has remained
to be open. 

In this section, we will develop two gradient parameter estimation schemes for the
discrete-time state feedback state tracking adaptive control problem:
a direct adaptive control scheme (see Section \ref{4Direct Adaptive
Control Design}) whose controller parameters are directly updated by
some adaptive laws, and an indirect adaptive control scheme (see
Section \ref{4Indirect Adaptive Control Design}) whose controller 
parameters are indirectly calculated from some adaptive parameter
estimates. 
\hspace*{\fill} $\fbox{}$
\end{rem}

\subsection{Direct Adaptive Control Design}
\label{4Direct Adaptive Control Design}
In this subsection, we develop the discrete-time direct adaptive
gradient state tracking control scheme, establish its desired
stability and tracking properties, and present an illustrative example.

\subsubsection{Adaptive Control Scheme}
For the adaptive control problem when the parameters $A$ and $b$ are
unknown (and so are $k_1^*$ and $k_2^*$), the nominal control law
(\ref{432A.4}) is replaced by its adaptive version:
\beq
u(t) = k^{T}_{1}(t) x(t) + k_{2}(t) r(t),
\label{432A.71}
\eeq
where $k_{1}(t)$ and $k_{2}(t)$ are the estimates of $k_{1}^{*}$ and
$k_{2}^{*}$, respectively. The adaptive control design task now is to
choose some desired adaptive laws (algorithms) to update $k_1(t)$ and $k_2(t)$ 
so that the stated control objective is still achievable in the
presence of the uncertainties of $A$ and $b$.

\medskip
\medskip
{\bf Error model}. Defining the parameter errors
\beq
\tilde{k}_{1}(t) = k_{1}(t) - k_{1}^{*},\;
\tilde{k}_{2}(t) = k_{2}(t) - k_{2}^{*}
\label{432A.8}
\eeq
and using (\ref{432A.1}), (\ref{4me}), and (\ref{432A.71}), we obtain the
closed-loop system
\bea
x(t+1) \ts = \ts A x(t) + b \left(k_{1}^{T}(t) x(t) 
+ k_{2}(t) r(t)\right) \nn \\
\ts = \ts A_{m} x(t) + b_{m} r(t) + b_{m} \left( \frac{1}{k_{2}^{*}} 
\tilde{k}_{1}^{T}(t) x(t) + \frac{1}{k_{2}^{*}} \tilde{k}_{2}(t)
r(t)\right).
\label{432A.9}
\eea
Substituting (\ref{432A.2}) in (\ref{432A.9}), we have the tracking error equation
\beq
e(t+1) = A_{m} e(t) +
b_{m} \frac{1}{k_{2}^{*}} \left(\tilde{k}_{1}^{T}(t) x(t)
+ \tilde{k}_{2}(t) r(t) \right).
\label{432A.10}
\eeq

Introducing $\rho^* = 1/k_2^*$ and 
\bea
\theta (t) \ts = \ts \left[k_1^T(t), k_2 (t)\right]^T \in R^{n+1}\\*[0.05in]
\theta^{*} \ts = \ts \left[k_1^{*T}, k_2^{*} \right]^T\in R^{n+1} \\*[0.05in]
\omega (t) \ts = \ts \left[x^{T}(t), r(t) \right]^{T}\in
R^{n+1} \label{4omega(t)} \\*[0.05in]
W_m(z) \ts = \ts (zI - A_m)^{-1} b_m = [w_{m1}(z), w_{m2}(z), \ldots, w_{mn}(z)]^T,
\eea
from (\ref{432A.10}), we obtain
\beq
e(t) = \rho^* W_m(z)[(\theta - \theta^*)^T \omega](t),\footnote{As a
  notation, $y(t) = G(z)[v](t)$ represents the output $y(t)$ of the
  system $G(z)$ with input $v(t)$.}
\label{432A.101}
\eeq
which, with $e(t) = [e_1(t), e_2(t), \ldots, e_n(t)]^T \in R^n$, is
equivalent to 
\beq
e_i(t) = \rho^* w_{mi}(z)[(\theta - \theta^*)^T
\omega](t),\;i=1,2,\ldots, n.
\label{432A.101a}
\eeq

Letting $\rho(t)$ be the estimate of $\rho^*$ and introducing
\bea
\zeta_i (t) \ts = \ts w_{mi}(z)[\omega](t)\in R^{n+1}\\*[0.05in]
\xi_i(t) \ts = \ts \theta^{T}(t) \zeta_i(t) - w_{mi} (z) [\theta^{T} \omega](t) \in R,
\label{4xii(t)}
\eea
we define the estimation errors
\beq
\epsilon_i (t) =  e_i(t) + \rho(t) \xi_i(t),\;i=1,2,\ldots, n,
\label{432A.35}
\eeq
and derive
\beq
\epsilon_i (t) = \rho^{*} (\theta(t) - \theta^{*})^{T} \zeta_i(t) + (\rho(t) - 
\rho^{*}) \xi_i(t),\;i=1,2,\ldots, n.
\label{432A.36}
\eeq

\medskip
{\bf Adaptive laws}. Similar to Assumption (A3), we make the following
assumption on $k_2^*$:

\begin{description}
\item[] {\bf Assumption (A4)}: A lower bound $k_2^a > 0$ of
$|k_2^*|$: $k_2^a \leq |k_2^*|$, is known.
\footnote{With $\rho^* = 1/k_2^*$, $\rho^b = 1/k_2^a$ is an upper bound of $|\rho^*|$:
$|\rho^*| \leq \rho^b$.}
\end{description}

Introducing the cost function 
$
J = \frac{1}{2} \frac{\sum_{i=1}^n \epsilon_i^2}{m^2}
$, 
where $m = m(t)$ is a normalization signal to be defined, 
and deriving its gradients
\bea
\frac{\partial J}{\partial \theta} \ts = \ts \rho^* \sum_{i=1}^n \frac{\epsilon_i
\zeta_i}{m^2} \\*[0.05in]
\frac{\partial J}{\partial \rho} \ts = \ts \sum_{i=1}^n \frac{\epsilon_i
\xi_i}{m^2},
\eea
we choose the adaptive laws for $\theta(t)$ and $\rho(t)$ as
\bea
\label{4thetat+1}
\theta(t+1) \ts =  \ts\theta(t) - \frac{\sign[\rho^*] \Gamma
\sum_{i=1}^n \epsilon_i(t) \zeta_i(t)}{m^2(t)} \\*[0.05in]
\rho(t+1) \ts = \ts \rho(t) - \frac{\gamma \sum_{i=1}^n \epsilon_i(t)
\xi_i(t)}{m^2(t)},
\label{4rhot+1}
\eea
where $0 < \Gamma = \Gamma^T < 2 k_2^a I_{n+1}$, $0 < \gamma < 2$, and
\beq
m(t) = \sqrt{1 + \sum_{i=1}^n \zeta_i^T(t) \zeta_i(t) + \sum_{i=1}^n
\xi_i^2(t)}.
\eeq

\subsubsection{Stability Analysis} 
Consider the positive definite function
\beq
V(\tilde{\theta}, \tilde{\rho}) = |\rho^{*}|\tilde{\theta}^{T} 
\Gamma^{-1} \tilde{\theta} + \gamma^{-1} \tilde{\rho}^{2},
\label{432A.40}
\eeq
where the parameter errors are
\beq
\tilde{\theta}(t) = \theta(t) - \theta^{*},\;
\tilde{\rho}(t) = \rho(t) - \rho^{*}.
\label{432A.41}
\eeq
The time-increment of $V(\tilde{\theta},\tilde{\rho})$, 
along the trajectories of (\ref{4thetat+1})-(\ref{4rhot+1}), is
\bea
\ts \ts V(\tilde{\theta}(t+1), \tilde{\rho}(t+1)) -
V(\tilde{\theta}(t), \tilde{\rho}(t)) \nn\\
\ts = \ts |\rho^*| \left(\tilde{\theta}(t) - \frac{\sign[\rho^*] \Gamma
\sum_{i=1}^n \epsilon_i(t) \zeta_i(t)}{m^2(t)}\right)^T \Gamma^{-1} 
\left(\tilde{\theta}(t) - \frac{\sign[\rho^*] \Gamma
\sum_{i=1}^n \epsilon_i(t) \zeta_i(t)}{m^2(t)}\right) \nn\\
\ts \ts + 
\left(\tilde{\rho}(t) - \frac{\gamma \sum_{i=1}^n \epsilon_i(t)
\xi_i(t)}{m^2(t)}\right) \gamma^{-1} \left(\tilde{\rho}(t) 
- \frac{\gamma \sum_{i=1}^n \epsilon_i(t)
\xi_i(t)}{m^2(t)}\right) \nn\\
\ts \ts - \left(|\rho^{*}|\tilde{\theta}^{T}(t)  
\Gamma^{-1} \tilde{\theta}(t) + \gamma^{-1} \tilde{\rho}^{2}(t)\right) \nn\\
\ts = \ts - \frac{2 \rho^* 
\sum_{i=1}^n \epsilon_i(t) \zeta_i^T(t) \tilde{\theta}(t)}{m^2(t)} + 
 \frac{\sum_{i=1}^n \epsilon_i(t) \zeta_i^T(t)}{m^2(t)}\, |\rho^*| \,\Gamma  \,
\frac{\sum_{i=1}^n \epsilon_i(t) \zeta_i(t)}{m^2(t)} \nn\\
\ts \ts - \frac{2 \sum_{i=1}^n \epsilon_i(t)
\xi_i(t) \tilde{\rho}(t)}{m^2(t)} + 
\frac{\sum_{i=1}^n \epsilon_i(t)
\xi_i(t)}{m^2(t)}\, \gamma\, \frac{\sum_{i=1}^n \epsilon_i(t)
\xi_i(t)}{m^2(t)} \nn\\
\ts = \ts - \frac{2 \sum_{i=1}^n \epsilon_i^2(t)}{m^2(t)} + 
 \frac{\sum_{i=1}^n \epsilon_i(t) \zeta_i^T(t)}{m^2(t)}\, |\rho^*| \,\Gamma  \,
\frac{\sum_{i=1}^n \epsilon_i(t) \zeta_i(t)}{m^2(t)} \nn\\
\ts \ts + \frac{\sum_{i=1}^n \epsilon_i(t)
\xi_i(t)}{m^2(t)}\, \gamma\, \frac{\sum_{i=1}^n \epsilon_i(t)
\xi_i(t)}{m^2(t)} 
\eea
where, with $\|\cdot\|_2$ being the $l^2$ vector norm and $\gamma_1
\in (0, 2)$ being the maximum eigenvalue of $|\rho^*| \,\Gamma$, 
\bea
\ts \ts \frac{\sum_{i=1}^n \epsilon_i(t) \zeta_i^T(t)}{m^2(t)}\, |\rho^*| \,\Gamma  \,
\frac{\sum_{i=1}^n \epsilon_i(t) \zeta_i(t)}{m^2(t)} \nn\\
\ts \leq \ts \gamma_1\, \frac{\sum_{i=1}^n |\epsilon_i(t)| \|\zeta_i(t)\|_2}{m^2(t)}\,
\frac{\sum_{i=1}^n |\epsilon_i(t)| \|\zeta_i(t)\|_2}{m^2(t)} \nn\\
\ts \leq \ts \gamma_1 \,\frac{\sqrt{\sum_{i=1}^n |\epsilon_i(t)|^2} 
\sqrt{\sum_{i=1}^n \|\zeta_i(t)\|_2^2}}{m^2(t)}\,
\frac{\sqrt{\sum_{i=1}^n |\epsilon_i(t)|^2} 
\sqrt{\sum_{i=1}^n \|\zeta_i(t)\|_2^2}}{m^2(t)} \nn\\
\ts = \ts \gamma_1 \,\frac{\sum_{i=1}^n |\epsilon_i(t)|^2 \sum_{i=1}^n
\|\zeta_i(t)\|_2^2}{m^4(t)} \nn\\
\ts = \ts 
\gamma_1 \,\frac{\sum_{i=1}^n \epsilon_i^2(t) \sum_{i=1}^n
\zeta_i^T(t) \zeta_i(t)}{m^4(t)},
\label{4ineq1}
\eea
where, Schwarz inequality is used for the second inequality above, and similarly, 
\bea
\ts \ts \frac{\sum_{i=1}^n \epsilon_i(t)
\xi_i(t)}{m^2(t)}\, \gamma\, \frac{\sum_{i=1}^n \epsilon_i(t)
\xi_i(t)}{m^2(t)} \nn\\
\ts \leq \ts \gamma \,\frac{\sum_{i=1}^n |\epsilon_i(t)|
|\xi_i(t)|}{m^2(t)}\, \frac{\sum_{i=1}^n |\epsilon_i(t)|
|\xi_i(t)|}{m^2(t)} \nn\\
\ts \leq \ts \gamma \,\frac{\sqrt{\sum_{i=1}^n |\epsilon_i(t)|^2} 
\sqrt{\sum_{i=1}^n |\xi_i(t)|^2}}{m^2(t)}
\, \frac{\sqrt{\sum_{i=1}^n |\epsilon_i(t)|^2} 
\sqrt{\sum_{i=1}^n |\xi_i(t)|^2}}{m^2(t)} \nn\\
\ts = \ts \gamma \, \frac{\sum_{i=1}^n |\epsilon_i(t)|^2 \sum_{i=1}^n
|\xi_i(t)|^2}{m^4(t)} \nn\\
\ts = \ts \gamma \, \frac{\sum_{i=1}^n \epsilon_i^2(t) \sum_{i=1}^n \xi_i^2(t)}{m^4(t)}.
\eea
Finally, for $\gamma_0 = \max\{\gamma_1, \gamma\} \in (0, 2)$, we arrive at
\beq
V(\tilde{\theta}(t+1), \tilde{\rho}(t+1)) -
V(\tilde{\theta}(t), \tilde{\rho}(t)) \leq - (2 - \gamma_0)
\frac{\sum_{i=1}^n \epsilon_i^2(t)}{m^2(t)} \leq 0,
\eeq
from which we have the following desired properties:

\begin{lem}
\label{4lemmadi}
The adaptive laws (\ref{4thetat+1})-(\ref{4rhot+1}) ensure:

\medskip
(i) $\theta(t)$, $\rho(t)$ and $\frac{\sum_{i=1}^n \epsilon_i^2(t)}{m^2(t)}$ are 
 bounded; and 

\medskip
(ii) $\frac{\epsilon_i(t)}{m(t)} \in L^2$, $i=1,2,\ldots, n$, $\theta(t+1)
- \theta(t) \in L^2$, $\rho(t+1) - \rho(t) \in L^{2}$.
\end{lem}

In this discrete-time case, Property (ii) implies: 
$\lim_{t \rightarrow \infty} \frac{\epsilon_i(t)}{m(t)} = 0$,
$i=1,2,\ldots, n$, $\lim_{t \rightarrow \infty} (\theta(t+1) - \theta(t)) = 0$, and
$\lim_{t \rightarrow \infty} (\rho(t+1) - \rho(t)) = 0$. We can then
establish the following result:

\begin{thm}
\label{4thm}
The adaptive controller (\ref{432A.71}), updated from the adaptive law
(\ref{4thetat+1})--(\ref{4rhot+1}) and applied to the plant
(\ref{432A.1}), ensures that all closed-loop system signals are bounded
and $\lim_{t \rightarrow \infty} (x(t) - x_m(t)) = 0$.
\end{thm}
{\bf Proof}: 
We first apply the discrete-time swapping lemma \cite[page
366]{if06}, \cite[page 411]{t03} to the signals $\xi_i(t)$ in
(\ref{4xii(t)}): for a minimal realization $(A_i, b_i, c_i)$ of
$w_{mi}(z) = c_i(zI - A)^{-1} b_i$, with $h_{ic}(z) = c_i(zI -
A)^{-1}$ and $h_{ib}(z) = (zI - A)^{-1} b$ both stable and strictly proper, $\xi_i(t)$ can be
expressed as
\bea
\label{4xii1}
\xi_i(t) \ts = \ts \theta^{T}(t) \zeta_i(t) - w_{mi} (z)
[\theta^{T} \omega](t) \nn\\
\ts = \ts h_{ic}(z)[(z-1)[\theta^T] z h_{ib}(z)[\omega]](t),
\eea
in which $(z-1)[\theta](t) = \theta(t+1)-\theta(t) \in L^2$ (that is,
$k_1(t+1) - k_1(t) \in L^2$, $k_2(t+1) - k_2(t) \in L^2$), 
$\omega(t) = \left[x^{T}(t), r(t) \right]^{T}$ and $z h_{ib}(z)$ is
stable and proper. 

The above $\xi_i(t)$ can be further expressed as
\bea
\xi_i(t) \ts = \ts h_{ic}(z)[(z-1)[\theta^T] h_{ib}(z)[z [\omega]]](t) \nn\\
\ts = \ts h_{ic}(z)[(z-1)[\theta^T]
h_{ib}(z)[[(A x + b (k_1^T x + k_2 r))^T, \bar{r}]^T]](t),
\label{4xii2}
\eea
where $\bar{r}(t) = r(t+1)$ and $z [\omega](t) = \omega(t+1) =
[x^T(t+1), r(t+1)]^T$ with $x(t+1) = A x(t) + b u(t)$ and $u(t) = k_1^T(t) x + k_2(t) r(t)$.

With $\epsilon(t) =
[\epsilon_1(t), \epsilon_2(t), \ldots, \epsilon_n(t)]^T$ and $\xi(t) =
[\xi_1(t), \xi_2(t), \ldots, \xi_n(t)]^T$, from (\ref{432A.35}), we
have
\beq
x(t) = x_m(t) + \epsilon(t) - \rho(t) \xi(t).
\label{4x(t)}
\eeq
Denoting the $l^1$ vector norm of $x(t)$ as 
\beq
\|x(t)\| = \|x(t)\|_1 = |x_1(t)| + |x_2(t)| + \cdots + |x_n(t)|,
\eeq
with $m(t) = \sqrt{1 
+ \sum_{i=1}^n \zeta_i^T(t) \zeta_i(t) + \sum_{i=1}^n \xi_i^2(t)}$,
from (\ref{4x(t)}), we obtain
\beq
\|x(t)\| \leq \|x_m(t)\| + \frac{\|\epsilon(t)\|}{m(t)} m(t) +
|\rho(t)| \|\xi(t)\|,
\label{4x(t)norm}
\eeq
where $\frac{\|\epsilon(t)\|}{m(t)} \in L^2 \cap L^\infty$, and 
\beq
m(t) \leq 1 + \sum_{i=1}^n \|\zeta_i(t)\| + \|\xi(t)\|.
\label{4m(t)ineq}
\eeq

\medskip
{\bf Basic operator concepts}. A linear time-invariant system has a
transfer function $T(z)$ which can be considered as an operator
operating on an input signal $u(t)$ to generate an output signal
$y(t)$: $y(t) = T(z)[u](t)$, where $z$ is the time-advance operator: $z[u](t) =
u(t+1)$, in discrete time. 

A (scalar) linear operator $T(z, t)$ represents a linear input-output
relationship of a possibly time-varying dynamic system whose output is 
$y(t) = T(z, \cdot)[u](t) \in R$, for an input $u(t) \in R$.

\begin{defn}
A linear operator $T(z, t)$ is stable and proper if
\beq
|y(t)| = |T(z, \cdot)[u](t)| \leq \beta \sum_{\tau = 0}^{t-1}
e^{-\alpha (t -1 - \tau)} |u(\tau)| + \gamma |u(t)|
\eeq
for any $u(t) \in R$, all $t \geq 0$, and some constants
$\beta > 0$, $\alpha > 0$, $\gamma > 0$. A linear operator 
$T(z, t)$ is stable and strictly proper if it is stable with $\gamma = 0$.
\end{defn}

\begin{prop}
A linear operator $T(z,t)$ is stable and proper if it represents a
system described by the difference equation
\beq
P(z)[y](t) = Q(z, t)[u](t),
\eeq
where $P(z)$ is an $n$th-order constant coefficient polynomial 
whose zeros are all inside the unit circle of the complex $z$-plane,
and $Q(z,t)$ is an $n$th-order polynomial with bounded and
possibly time-varying coefficients. If the order of $Q(z, t)$ is less
than $n$, then $T(z,t)$ is stable and strictly proper.
\end{prop}

\begin{defn}
A linear operator $T(z,t)$ is nonnegative if
$T(z, \cdot)[u](t) \geq 0$, $\forall u(t) \geq 0$, $\forall t \geq 0$. 

A nonnegative linear operator $T_1(z,t)$ dominates a
 linear operator $T_2(z,t)$ if 
\beq
|T_2(z, \cdot)[u](t)| \leq T_1(z, \cdot)[u](t),\;\forall u(t) \geq
0,\;\forall t \geq 0. 
\eeq

A nonnegative linear operator $T(z,t)$ is nondecreasing if 
\beq
T(z,t)[u_1](t) \leq T(z,t)[u_2](t),\;\forall u_2(t) \geq
u_1(t) \geq 0,\; \forall t \geq 0.
\eeq
\end{defn}

\begin{prop}
For any stable and proper (strictly proper) linear operator $T_2(z,
t)$, there exists a nonnegative, stable and proper (strictly proper) linear
operator $T_1(z,t)$ which dominates $T_2(z,t)$. Such an operator
$T_1(z,t)$ can be chosen to be nondecreasing.
\end{prop}

{\bf Operator-based signal analysis}. For $\zeta_i(t) =
w_{mi}(z)[\omega](t)$ in (\ref{4m(t)ineq}) with $\omega(t)
= \left[x^{T}(t), r(t) \right]^{T}$, there exists a stable, 
strictly proper and nonnegative operator $T_{\zeta_i}(z)$ such that
\beq
\|\zeta_i(t)\| \leq T_{\zeta_i}(z)[\|x\|](t) + c_{\zeta_i},
\label{4zetain}
\eeq
for some constant $c_{\zeta_i} > 0$. 
For $\xi_i(t) = \theta^{T}(t) \zeta_i(t) - w_{mi} (z)
[\theta^{T} \omega](t)$ in (\ref{4xii1}), there exist a stable, 
strictly proper and nonnegative operator $T_{\xi_i}(z)$ and a constant $c_{\xi_i}
> 0$ such that 
\beq
\|\xi_i(t)\| \leq T_{\xi_i}(z)[\|x\|](t) + c_{\xi_i}.
\eeq
With $\xi(t) = [\xi_1(t), \xi_2(t), \ldots, \xi_n(t)]^T$ in
(\ref{4m(t)ineq}), there exist a stable, 
strictly proper and nonnegative operator $T_{m}(z)$ and a constant $c_{m}
> 0$ such that 
\beq
\sum_{i=1}^n \|\zeta_i(t)\| + \|\xi(t)\| \leq T_{m}(z)[\|x\|](t) + c_{m}.
\eeq

For $\xi_i(t) = h_{ic}(z)[(z-1)[\theta^T]
h_{ib}(z)[[(A x + b (k_1^T x + k_2 r))^T, \bar{r}]^T]](t)$ in
(\ref{4xii2}), for $\Delta_{k_1}(t) = k_1(t+1) - k_1(t) \in L^2$, 
there exist stable, strictly proper and nonnegative
operators $T_{ic}(z)$ and $T_{ib}(z, t)$ and a constant $c_i > 0$ such
that
\beq
|\xi_i(t)| \leq T_{ic}(z)[\|\Delta_{k_1}\|\, T_{ib}(z, \cdot)[\|x\|]](t) +
 c_i,
\eeq
and for the corresponding vector $\xi(t) =
[\xi_1(t), \xi_2(t), \ldots, \xi_n(t)]^T$, we have
\beq
\|\xi(t)\| \leq T_{c}(z)[\|\Delta_{k_1}\|\, T_{b}(z, \cdot)[\|x\|]](t) +
c_{\xi}
\label{4xinorm}
\eeq
for some stable, strictly proper and nonnegative
operators $T_{c}(z)$ and $T_{b}(z, t)$ and some constant $c_{\xi} >
0$. From (\ref{4x(t)norm}) and (\ref{4zetain})-(\ref{4xinorm}), we obtain
\bea
\|x(t)\| \ts \leq \ts \|x_m(t)\| + \frac{\|\epsilon(t)\|}{m(t)}\big(1 +
T_{m}(z)[\|x\|](t) + c_{m}\big) + |\rho(t)| 
\big(T_{c}(z)[\|\Delta_{k_1}\|\, T_{b}(z, \cdot)[\|x\|]](t) +
c_{\xi}\big) \nn\\
\ts = \ts \frac{\|\epsilon(t)\|}{m(t)} T_{m}(z)[\|x\|](t) + |\rho(t)| 
T_{c}(z)[\|\Delta_{k_1}\|\, T_{b}(z, \cdot)[\|x\|]](t) \nn\\
\ts \ts + \|x_m(t)\| + \frac{\|\epsilon(t)\|}{m(t)}(1 + c_m) +
|\rho(t)| c_{\xi},
\label{4x(t)norm1}
\eea
where $\|x_m(t)\| + \frac{\|\epsilon(t)\|}{m(t)}(1 + c_m) + |\rho(t)|
c_{\xi}$ is bounded. There exists a stable, strictly proper,
nonnegative and nondecreasing operator $T_0(z)$ such that 
\bea
\ts \ts \frac{\|\epsilon(t)\|}{m(t)} T_{m}(z)[\|x\|](t) + |\rho(t)| 
T_{c}(z)[\|\Delta_{k_1}\|\, T_{b}(z, \cdot)[\|x\|]](t) \nn\\
\ts \leq \ts \frac{\|\epsilon(t)\|}{m(t)} T_{0}(z)[\|x\|](t) + |\rho(t)| 
T_{c}(z)[\|\Delta_{k_1}\|\, T_{0}(z)[\|x\|]](t).
\label{4ineq11}
\eea
From (\ref{4x(t)norm1}) and (\ref{4ineq11}), it follows that 
\bea
T_0(z)[\|x\|](t) \ts \leq \ts T_0(z)[\frac{\|\epsilon\|}{m}
T_{0}(z)[\|x\|]](t)  + T_0(z)[|\rho| T_{c}(z)[\|\Delta_{k_1}\|\, T_{0}(z)[\|x\|]]](t)
+ c_0, 
\label{4ineq2}
\eea
where $c_0 \geq T_0(z)[\|x_m\| + \frac{\|\epsilon\|}{m}(1 + c_m) + |\rho|
c_{\xi}](t)$ is a constant. For (\ref{4ineq2}), there exists a stable
and strictly proper operator $T(z)$ such that
\bea
T_0(z)[\|x\|](t) \ts \leq \ts T(z)[\frac{\|\epsilon\|}{m}
T_{0}(z)[\|x\|]](t) + T(z)[\|\Delta_{k_1}\|\, T_{0}(z)[\|x\|]](t) + c_0 \nn\\
\ts = \ts T(z)[\left(\frac{\|\epsilon\|}{m} + \|\Delta_{k_1}\|\right)
T_{0}(z)[\|x\|]](t) + c_0,
\label{4ineq3}
\eea
where $\frac{\|\epsilon(t)\|}{m}(t) + \|\Delta_{k_1}(t)\| \in L^2 \cap
L^\infty$. The $L^2$ property of $\frac{\|\epsilon(t)\|}{m}(t)
+ \|\Delta_{k_1}(t)\|$ ensures a small gain for the feedback structure
in terms of $T_0(z)[\|x\|](t)$ in (\ref{4ineq3}). A small gain theorem
can be applied to (\ref{4ineq3}), to prove that $T_0(z)[\|x\|](t)$ is
bounded, and so is $\|x(t)\|$ from (\ref{4x(t)norm1}) and
(\ref{4ineq11}), so that $u(t) = k_1^T(t) x + k_2(t) r(t)$ is
bounded. Thus, all system signals are bounded.

Then, in (\ref{4x(t)}), $\epsilon(t) \in L^2$ and $\xi(t) \in
L^2$ as $\xi_i(t) \in L^2$ from (\ref{4xii2}) with $(z-1)[\theta](t) = \theta(t+1)
- \theta(t) \in L^2$. Finally, $x(t) - x_m(t) \in L^2$ so that
$\lim_{t \rightarrow \infty} (x(t) - x_m(t)) = 0$.
\hspace*{\fill} $\nabla$ 

\bigskip
The above adaptive control scheme is called a direct adaptive control
scheme, as it directly updates the controller parameters $\theta (t) =
[k_1^T(t), k_2 (t)]^T$.

\subsubsection{An Illustrative Example}
Consider the second-order plant:
\beq
\left[
\begin{array}{c}
x_1(t+1)\\
x_2(t+1)
\end{array}
\right] = 
\left[
\begin{array}{cc}
1 & -1\\
2 & 1
\end{array}
\right] 
\left[
\begin{array}{c}
x_1(t)\\
x_2(t)
\end{array}
\right] + 
\left[
\begin{array}{c}
0\\
2
\end{array}
\right] u(t),
\eeq
with $\det(zI - A) = z^2 - 2 z + 3$ and poles: $-1 \pm \sqrt{2}\,i$ ($i=\sqrt{-1}$), and 
the reference model system:
\beq
\left[
\begin{array}{c}
x_{m1}(t+1)\\
x_{m2}(t+1)
\end{array}
\right] = 
\left[
\begin{array}{cc}
1 & -1\\
1.05 & -1.2
\end{array}
\right] 
\left[
\begin{array}{c}
x_{m1}(t)\\
x_{m2}(t)
\end{array}
\right] + 
\left[
\begin{array}{c}
0\\
1
\end{array}
\right] r(t),
\eeq
whose transfer matrix (vector) is
\bea
W_m(z) \ts = \ts (z I - A_m)^{-1} b_m = [w_{m1}(z), w_{m2}(z)]^T \nn\\
\ts = \ts 
[\frac{-1}{z^2+0.2z-0.15}, \frac{z-1}{z^2+0.2z-0.15}]^T
\eea
 with poles: $0.3$ and $-0.5$. The matching parameters $k_1^*$ and
 $k_2^*$ for (\ref{4me}) are
\beq
k_1^* = [-0.475, -1.1]^T,\;k_2^* = 0.5.
\eeq
With $\omega(t) = [x^T(t), r(t)]^T$ and $\theta(t) = [k_1^T(t),
k_2(t)]^T$, we have
\beq
\zeta_i(t) = w_{mi}(z)[\omega](t),\;i=1,2,
\eeq
and with $u(t) = \theta^T(t) \omega(t)$, we have
\beq
\xi_i(t) = \theta^T(t) \zeta_i(t) - w_{mi}(z)[u](t),\;i=1,2.
\eeq
Then, we generate the estimation errors (\ref{432A.35}):
\beq
\epsilon_i (t) =  e_i(t) + \rho(t) \xi_i(t),\;i=1,2, 
\eeq
where $\rho(t)$ is the estimate of $\rho^{*} =1/k_2^* = 2$. 

The adaptive laws for $\theta(t)$ and $\rho(t)$ are from 
(\ref{4thetat+1})-(\ref{4rhot+1}):
\bea
\label{4theta(t+1)}
\theta(t+1) \ts =  \ts\theta(t) - \frac{\sign[\rho^*] \Gamma
\sum_{i=1}^n \epsilon_i(t) \zeta_i(t)}{m^2(t)} \\*[0.05in]
\rho(t+1) \ts = \ts \rho(t) - \frac{\gamma \sum_{i=1}^n \epsilon_i(t)
\xi_i(t)}{m^2(t)},
\eea
where $0 < \Gamma = \Gamma^T < 2 k_2^a I_{3}$ with $k_2^a \leq |k_2^*|
= 0.5$, $0 < \gamma < 2$, and
\beq
m^2(t) = 1 + \sum_{i=1}^2 \zeta_i^T(t) \zeta_i(t) + \sum_{i=1}^2
\xi_i^2(t).
\eeq
The control law is from (\ref{432A.71}):
\beq
u(t) = k^{T}_{1}(t) x(t) + k_{2}(t) r(t),
\eeq
where $k_{1}(t) \in R^2$ and $k_{2}(t) \in R$ are from $\theta(t)$: 
$\theta (t) = \left[k_1^T(t), k_2 (t)\right]^T \in R^{3}$, as the
adaptive estimate of the unknown parameter vector $\theta^* =
[k_1^{*T}, k_2^*]^T$, obtained from (\ref{4theta(t+1)}).

\bigskip
{\bf Simulation results}. 
The simulation results are shown in Figure 1 (the plant state
$x_1(t)$, reference model state $x_{m1}(t)$, and tracking error
$e_1(t) = x_1(t) - x_{m1}(t)$), 
for $r(t) = \sin (0.13t)$, $\Gamma = 0.5
I_3$, $\gamma = 1.5$, $\theta(0) = 1.25 \theta^*$ and $\rho(0) =
1.25 \rho^*$, and in Figure 2 (the plant state
$x_2(t)$, reference model state $x_{m2}(t)$, and tracking error 
$e_2(t) = x_2(t) - x_{m2}(t)$), for the same conditions. 

Another set of simulation results are shown in Figure 3 (the plant state
$x_1(t)$, reference model state $x_{m1}(t)$, and tracking error
$e_1(t) = x_1(t) - x_{m1}(t)$), 
for $r(t) = \sin (0.13t) + \sin(1.3t)$, $\Gamma = 0.5
I_3$, $\gamma = 1.5$, $\theta(0) = 1.25 \theta^*$ and $\rho(0) =
1.25 \rho^*$, and in Figure 4 (the plant state
$x_2(t)$, reference model state $x_{m2}(t)$, and tracking error 
$e_2(t) = x_2(t) - x_{m2}(t)$), for the same conditions. 

\begin{figure}[p]
\centering
\includegraphics[height=4in,width=5.0in,angle=0]{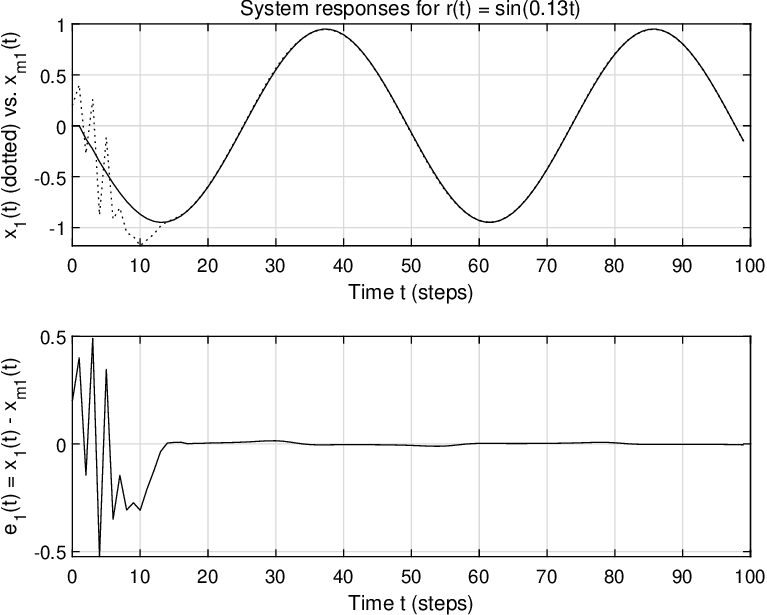}
\caption{System responses: $x_1(t), x_{m1}(t), e_1(t)$, for $r(t) = \sin(0.13t)$.}
\bigskip
\bigskip
\centering
\includegraphics[height=4in,width=5.0in,angle=0]{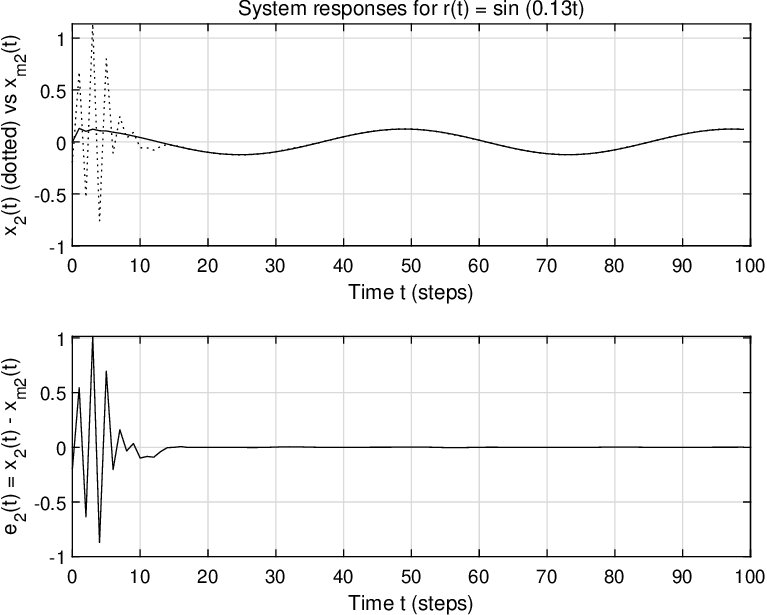}
\caption{System responses: $x_2(t), x_{m2}(t), e_2(t)$, for $r(t) = \sin(0.13t)$.}
\end{figure}

\begin{figure}[p]
\centering
\includegraphics[height=4in,width=5.0in,angle=0]{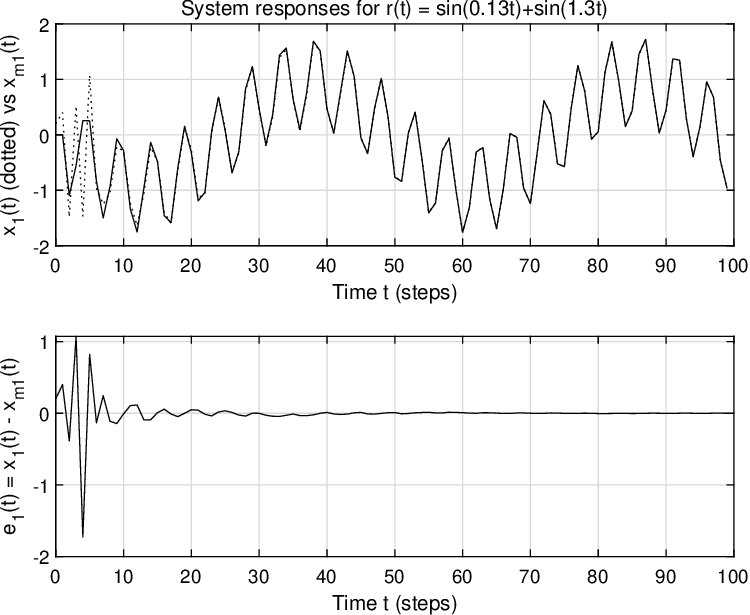}
\caption{Responses: $x_1(t), x_{m1}(t), e_1(t)$, for $r(t)
= \sin(0.13t) + \sin(1.3t)$.}
\bigskip
\bigskip
\centering
\includegraphics[height=4in,width=5.0in,angle=0]{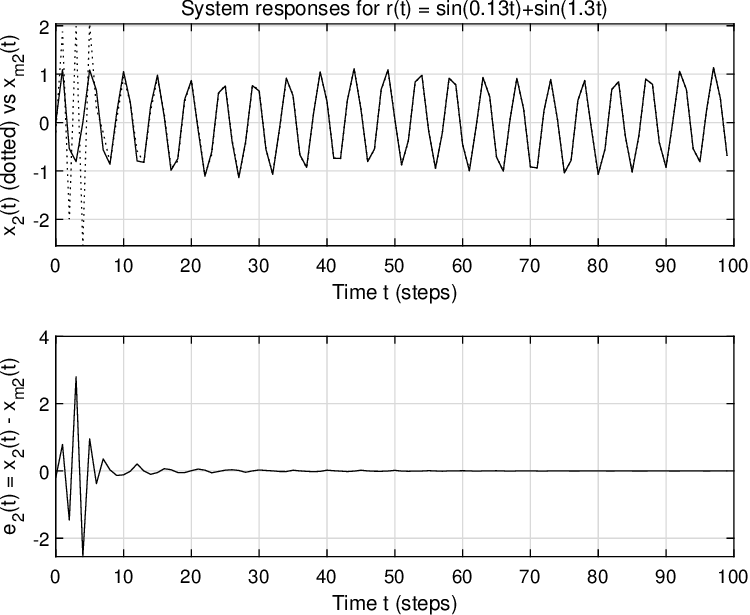}
\caption{Responses: $x_2(t), x_{m2}(t), e_2(t)$, for $r(t)
= \sin(0.13t) + \sin(1.3t)$.}
\end{figure}

\subsubsection{Application to Continuous-Time Systems}
The gradient design and analysis procedure
developed in this subsection for 
the discrete-time case is also applicable to the continuous-time case,
with those discrete-time transfer functions replaced by their
corresponding counterpart continuous-time transfer functions, and with
a continuous-time system analysis method (using the time-derivative of
$V(\tilde{\theta}, \tilde{\rho}) = 
|\rho^{*}|\tilde{\theta}^{T} 
\Gamma^{-1} \tilde{\theta} + \gamma^{-1} \tilde{\rho}^{2}$ as in 
(\ref{432A.40})) for the continuous-time versions of
the adaptive laws (\ref{4thetat+1})-(\ref{4rhot+1}), given by
\bea
\label{4thetat+1c}
\dot{\theta}(t) \ts = \ts - \frac{\sign[\rho^*] \Gamma
\sum_{i=1}^n \epsilon_i(t) \zeta_i(t)}{m^2(t)},\;\Gamma = \Gamma^T > 0 \\*[0.05in]
\dot{\rho}(t) \ts = \ts - \frac{\gamma \sum_{i=1}^n \epsilon_i(t)
\xi_i(t)}{m^2(t)},\;\gamma > 0.
\label{4rhot+1c}
\eea

\begin{rem}
\rm
The Lyapunov design and analysis procedure for the continuous-time
case consists of (\ref{4plantcc})-(\ref{4Vdotc}) and the control law
(\ref{432A.7}), which may not be applicable to the discrete-time case.

\medskip
To further illustrate this situation, from (\ref{432A.10})-(\ref{4omega(t)}), we have
\beq
e(t+1) = A_{m} e(t) +
b_{m} \frac{1}{k_{2}^{*}} \tilde{\theta}^{T}(t) \omega(t),
\label{432A.10a}
\eeq
where $\tilde{\theta}(t) = \theta(t) - \theta^*$. Consider the positive definite function
\beq
V(e, \tilde{\theta}) = e^T P e + \frac{1}{|k_2^*|} \tilde{\theta}^T \Gamma^{-1} \tilde{\theta},
\eeq
where $\Gamma = \Gamma^T > 0$ is to be further specified, 
$P = P^T > 0$ satisfying $A_m^T P A_m - P = - Q$ for a chosen
$Q = Q^T > 0$, take the adaptive law for $\theta(t)$ in the form
\beq
\theta(t+1) = \theta(t) + g(t),
\eeq
where $g(t) \in R^{n+1}$ is to be selected, and obtain the time-increment of $V$ as
\bea
\Delta V \ts = \ts V(e(t+1), \tilde{\theta}(t+1)) -
V(e(t), \tilde{\theta}(t)) \nn\\
\ts = \ts - e^T(t) Q e(t) + 
2 \omega^T(t) \tilde{\theta}(t) \frac{1}{k_2^*} b_m^T P A_m e(t) 
 + \omega^T(t) \tilde{\theta}(t) \frac{1}{k_2^*} b_m^T P
b_m \frac{1}{k_2^*} \tilde{\theta}^T(t) \omega(t) \nn\\
\ts \ts + 
\frac{2}{|k_2^*|} g^T(t) \Gamma^{-1} \tilde{\theta}(t) + 
\frac{1}{|k_2^*|} g^T(t) \Gamma^{-1} g(t).
\eea

It would be desirable if $g(t)$ could make $\Delta V \leq -
e^T(t) Q e(t)$ (similar to (\ref{4Vdotc})), for desired system
properties (similar to that in Proposition \ref{4prop}), but
such a $g(t)$ is yet to be derived, especially, the right sides of 
(\ref{4k1dotc})-(\ref{4k2dotc}) as the components of $g(t)$ cannot meet
such a need. \hspace*{\fill} $\fbox{}$
\end{rem}

\subsection{Indirect Adaptive Control Design}
\label{4Indirect Adaptive Control Design}
Consider the discrete-time SIMO linear time-invariant plant (\ref{432A.1}):
\beq
x(t+1) = A x(t) + b u(t),\;
x(t) \in R^{n},\;u(t) \in R
\label{432A.1-1}
\eeq
and the reference model system (\ref{432A.2}):
\beq
x_{m}(t+1) = A_{m} x_{m}(t) + b_{m} r(t),\;
x_{m}(t) \in R^{n},\;r(t) \in R
\label{432A.2-1}
\eeq
under the matching condition (\ref{4me}):
\beq
A + b k_{1}^{*T} = A_{m},\;
b k_{2}^{*} = b_{m}.
\label{432A.3-1}
\eeq

\medskip
{\bf Plant parametrization}. From the above matching condition, we express
\beq
A = A_m - b_m \theta_1^{*T},\;b = b_m \theta_2^*,
\eeq
where $\theta_1^* = k_2^{*-1} k_1^*$ and $\theta_2^* =
k_2^{*-1}$. This expression shows that the parameter uncertainties of
$A$ and $b$ are essentially that of $\theta_1^*$ and
$\theta_2^*$. Hence, the parameters $\theta_1^*$ and
$\theta_2^*$ can be considered as the unknown parts of the plant 
parameters $A$ and $b$, and we can parametrize the plant (\ref{432A.1-1}) as
\bea
x(t+1) \ts = \ts A x(t) + b u(t) = A_m x(t) + b_m (\theta_2^{*} u(t) -
\theta_1^{*T} x(t)).
\label{432para}
\eea

\medskip
{\bf Estimator parametrization}. We now design an adaptive parameter
estimation algorithm to estimate the unknown parameters $\theta_1^*$
and $\theta_2^*$ for adaptive control.

Letting $\theta_1(t)$ and $\theta_2(t)$ be the
estimates of $\theta_1^*$ and $\theta_2^*$, based on (\ref{432para}), we
 construct an adaptive {\it a posteriori} state estimator described by
 the dynamic equation
\beq 
\hat{x}(t+1) = A_m \hat{x}(t) + b_m
(\theta_2(t) u(t) - \theta_1^{T}(t) x(t)), 
\label{432est} 
\eeq 
to generate its state vector $\hat{x}(t)$ as an adaptive {\it a
posteriori} estimate of the system state $x(t)$. 
For the state estimation error $e_x(t) = \hat{x}(t) - x(t)$, from
(\ref{432para})-(\ref{432est}), we have the
state estimation error dynamic system equation 
\beq 
e_x(t+1) = A_m e_x(t) + b_m((\theta_2(t) - \theta_2^*) u(t) -
(\theta_1(t) - \theta_1^*)^T x(t)), 
\label{432eee}
\eeq
of the similar form as that in (\ref{432A.10}) without $1/k_2^*$: 
\beq
e(t+1) = A_{m} e(t) + b_{m} \left((k_{1}(t) - k_1^*)^T x(t) +
(k_{2}(t) - k_2^*) r(t)\right).
\eeq

Hence, (\ref{432eee}) can be expressed in the form
of (\ref{432A.101}) with $\rho^* = 1$ as
\beq
e_x(t) = W_m(z)[(\theta - \theta^*)^T \omega](t),
\label{432A.102}
\eeq
where
\bea
\theta (t) \ts = \ts \left[\theta_1^T(t), \theta_2 (t)\right]^T \in R^{n+1}\\*[0.05in]
\theta^{*} \ts = \ts \left[\theta_1^{*T}, \theta_2^{*} \right]^T\in R^{n+1} \\*[0.05in]
\omega (t) \ts = \ts \left[-x^{T}(t), u(t) \right]^{T}\in
R^{n+1}\\*[0.05in]
W_m(z) \ts = \ts (zI - A_m)^{-1} b_m = [w_{m1}(z), w_{m2}(z), \ldots, w_{mn}(z)]^T.
\eea

Based on the state estimation error equation (\ref{432A.102}), we can
design an adaptive scheme to update the parameter estimate
 $\theta(t)$, using a similar procedure to that for the
direct adaptive control case.

\medskip
With $e_x(t) = [e_{x1}(t), e_{x2}(t), \ldots, e_{xn}(t)]^T \in R^n$, we write
(\ref{432A.102}) as 
\beq
e_{xi}(t) = w_{mi}(z)[(\theta - \theta^*)^T
\omega](t),\;i=1,2,\ldots, n,
\eeq
and define the estimation errors
\beq
\epsilon_i (t) =  e_{xi}(t) + \xi_i(t),\;i=1,2,\ldots, n,
\label{432A.351}
\eeq
where
\bea
\xi_i(t) \ts = \ts \theta^{T}(t) \zeta_i(t) - W_{mi} (z)
[\theta^{T} \omega](t) \in R \\*[0.05in]
\zeta_i (t) \ts = \ts w_{mi}(z)[\omega](t)\in R^{n+1},
\eea
and derive the estimation error equations
\beq
\epsilon_i (t) = (\theta(t) - \theta^{*})^{T} \zeta_i(t),\;i=1,2,\ldots, n.
\label{432A.361}
\eeq

\medskip
{\bf Adaptive laws}. We choose the adaptive law for $\theta(t)$ as
\bea
\label{4thetat+11}
\theta(t+1) \ts =  \ts\theta(t) - \frac{\Gamma
\sum_{i=1}^n \epsilon_i(t) \zeta_i(t)}{m^2(t)} + f_\theta(t),
\eea
where $\Gamma = \diag\{\Gamma_1, \gamma_2\}$ with $\Gamma_1 \in R^{n
\times n}$ and $\gamma_2 \in R$ such that $0 < \Gamma_1 = \Gamma_1^T <
2 I_{n}$ and $\gamma_2 \in (0, 2)$, 
\beq
m(t) = \sqrt{1 + \sum_{i=1}^n \zeta_i^T(t) \zeta_i(t)}
\eeq
is the normalizing signal, 
and 
\beq
f_\theta(t) = [0, 0, \ldots, 0, f_2(t)]^T \in R^{n+1}
\eeq
with $f_2(t) \in R$ being a projection signal for $\theta_2(t)$ in
$\theta(t) = [\theta_1^T(t), \theta_2(t)]^T$, to ensure: 
\beq
\sign[\theta_2(t)] = \sign[\theta_2^*] = \sign[k_2^*]
\label{4theta2-1}
\eeq
\beq
|\theta_2(t)| \geq \theta_2^a,
\label{4theta2-2}
\eeq
where $\theta_2^a > 0$ is such that $|\theta_2^*| \geq \theta_2^a$ for
$\theta_2^* = 1/k_2^*$ ($\theta_2^a = 1/k_2^b$, see Assumption (A3)).

\bigskip
{\bf Parameter projection}. For $g_2(t) \in R$ being the last component of 
\beq
g(t) = - \frac{\Gamma
\sum_{i=1}^n \epsilon_i(t) \zeta_i(t)}{m^2(t)} = [g_1^T(t), g_2(t)]^T
\in R^{n+1}
\eeq
in (\ref{4thetat+11}), corresponding to $\theta(t) =
[\theta_1^T(t), \theta_2(t)]^T$, we choose
$\theta_2(0)$ such that $\sign[\theta_2(0)] = \sign[\theta_2^*] =
\sign[k_2^*]$ and $|\theta_2(0)| \geq \theta_2^a$, and set 
\beq
f_2(t) = \left\{ \begin{array}{ll}
0 & \mbox{if $\sign[k_2^*](\theta_2(t)+g_2(t)) \geq \theta_2^a$, }\\
\theta_2^a - \theta_2(t) - g_2(t) & \mbox{otherwise,}
\end{array}
\right.
\label{432proj}
\eeq
to satisfy (\ref{4theta2-1})-(\ref{4theta2-2}), as well as 
\beq
(\theta_2(t) - \theta_2^* + g_2(t) + f_2(t)) f_2(t) \leq 0.
\label{4125pp1d}
\eeq

\medskip
{\bf Stability analysis}. Consider the positive definite function
\beq
V(\tilde{\theta}) = \tilde{\theta}^{T} 
\Gamma^{-1} \tilde{\theta},\;\tilde{\theta}(t) = \theta(t) -
\theta^{*}. 
\label{432A.40in}
\eeq
The time-increment of $V(\tilde{\theta})$, 
along the trajectories of (\ref{4thetat+11}) without $f_\theta(t)$, is
\bea
\ts \ts V(\tilde{\theta}(t+1)) -V(\tilde{\theta}(t)) \nn\\
\ts = \ts \left(\tilde{\theta}(t) - \frac{\Gamma
\sum_{i=1}^n \epsilon_i(t) \zeta_i(t)}{m^2(t)}\right)^T \Gamma^{-1} 
\left(\tilde{\theta}(t) - \frac{\Gamma
\sum_{i=1}^n \epsilon_i(t) \zeta_i(t)}{m^2(t)}\right) \nn\\
\ts \ts - \tilde{\theta}^{T}(t) \Gamma^{-1} \tilde{\theta}(t)\nn\\
\ts = \ts - \frac{2 \sum_{i=1}^n \epsilon_i(t) \zeta_i^T(t) \tilde{\theta}(t)}{m^2(t)} + 
 \frac{\sum_{i=1}^n \epsilon_i(t) \zeta_i^T(t)}{m^2(t)}\, \Gamma  \,
\frac{\sum_{i=1}^n \epsilon_i(t) \zeta_i(t)}{m^2(t)} \nn\\
\ts = \ts - \frac{2 \sum_{i=1}^n \epsilon_i^2(t)}{m^2(t)} + 
 \frac{\sum_{i=1}^n \epsilon_i(t) \zeta_i^T(t)}{m^2(t)}\, \Gamma  \,
\frac{\sum_{i=1}^n \epsilon_i(t) \zeta_i(t)}{m^2(t)}
\eea
where, with $\|\cdot\|_2$ being the $l^2$ vector norm and $\gamma_1
\in (0, 2)$ being the maximum eigenvalue of $\Gamma$, and similar to (\ref
{4ineq1}), 
\beq
\frac{\sum_{i=1}^n \epsilon_i(t) \zeta_i^T(t)}{m^2(t)}\,\Gamma  \,
\frac{\sum_{i=1}^n \epsilon_i(t) \zeta_i(t)}{m^2(t)} \leq 
\gamma_1 \,\frac{\sum_{i=1}^n \epsilon_i^2(t) \sum_{i=1}^n
\zeta_i^T(t) \zeta_i(t)}{m^4(t)}.
\eeq
It follows that 
\beq
V(\tilde{\theta}(t+1)) - V(\tilde{\theta}(t)) \leq - (2 - \gamma_1)
\frac{\sum_{i=1}^n \epsilon_i^2(t)}{m^2(t)} \leq 0.
\label{4VDelta}
\eeq
Based on its key property (\ref{4125pp1d}), the parameter projection
signal $f_\theta(t)$ only adds some nonpositive term in 
$V(\tilde{\theta}(t+1)) - V(\tilde{\theta}(t))$, so that from
(\ref{4VDelta}), we can derive the following results:

\begin{lem}
\label{4lemmadiin}
The adaptive law (\ref{4thetat+11}) ensures:

\medskip
(i) $\theta(t)$ and $\frac{\sum_{i=1}^n \epsilon_i^2(t)}{m^2(t)}$ are 
 bounded; and 

\medskip
(ii) $\frac{\sum_{i=1}^n \epsilon_i^2(t)}{m^2(t)} \in L^1$, that is, 
$\frac{\epsilon_i(t)}{m(t)} \in L^2$, $i=1,2,\ldots, n$, 
$\theta(t+1) - \theta(t) \in L^{2}$.
\end{lem}

\medskip
\medskip
{\bf Control signal}. With $\theta(t) =
[\theta_1^T(t), \theta_2(t)]^T$, we design the adaptive control signal
 as
\beq
u(t) = k_1^T(t) x(t) + k_2(t) r(t),\;
k_1(t) = \frac{\theta_1(t)}{\theta_2(t)},\;k_2(t) = \frac{1}{\theta_2(t)}.
\label{432acl}
\eeq
The adaptive law (\ref{4thetat+11}) with parameter projection on
$\theta_2(t)$ ensures $|\theta_2(t)| \geq \theta_2^a > 0$ (that is,
$\sign[\theta_2^*] \theta_2(t) \geq \theta_2^a > 0$) for all $t \geq
0$, so that the adaptive control law (\ref{432acl}) is implementable.

\medskip
This is an indirect adaptive control design, as it first
updates the estimates $\theta(t) = [\theta_1^T(t), \theta_2 (t)]^T$ of
the uncertain plant parameters $\theta^{*} = [\theta_1^{*T},
  \theta_2^{*}]^T$ in (\ref{432para}), and then calculates the
controller parameters $k_1(t)$ and $k_2(t)$ from the plant parameter
estimates $\theta(t)$. 

Similar to that in Theorem \ref{4thm}, the closed-loop signal
boundedness and asymptotic state tracking ($\lim_{t \rightarrow
  \infty} (x(t) - x_m(t)) = 0$) are also ensured by this adaptive
control scheme. 

To see this, we note that, with the control law (\ref{432acl}), the estimator
equation (\ref{432est}) becomes
\beq 
\hat{x}(t+1) = A_m \hat{x}(t) + b_m r(t),
\eeq  
which, compared with the reference system equation (\ref{432A.2-1}),
implies that $\hat{x}(t) = x_m(t)$, so that (\ref{432A.351}) also has
the form (\ref{4x(t)}) with $\rho(t) = 1$ and the proof of 
Theorem \ref{4thm} for the direct adaptive control scheme is then 
also applicable to this indirect adaptive control scheme.

\begin{rem}
\rm
Such an indirect adaptive control design using a gradient algorithm
based adaptive law is also applicable to the continuous-time case,
based on the error equation (\ref{4eee}) which has a similar form to
that in (\ref{432eee}), with $\dot{e}_x(t)$ replacing $e_x(t+1)$, and
with a continuous-time version of the adaptive law (\ref{4thetat+11}): 
\beq
\dot{\theta}(t) = - \frac{\Gamma \sum_{i=1}^n \epsilon_i(t)
  \zeta_i(t)}{m^2(t)} + f_\theta(t),
\eeq
where $\Gamma = \diag\{\Gamma_1, \gamma_2\}$ with $\Gamma_1 \in R^{n
\times n}$ and $\gamma_2 \in R$ such that $\Gamma_1 = \Gamma_1^T > 0$
and $\gamma_2 > 0$, $m(t) = \sqrt{1 + \sum_{i=1}^n \zeta_i^T(t)
  \zeta_i(t)}$, and $f_\theta(t) = [0, \ldots, 0, f_2(t)]^T$ is the
projection signal whose component $f_2(t) \in R$ can be similarly
designed to that in (\ref{4f2(t)}). \hspace*{\fill} $\fbox{}$
\end{rem}

\setcounter{equation}{0}
\section{Designs for Multi-Input Multi-Output Systems}
We now consider a multi-input multi-output (MIMO) linear time-invariant plant
\beq
\left.
\begin{array}{c}
\dot{x}(t) \\
x(t+1)
\end{array}
\right\} 
= A x(t) + B u(t),
\label{42.2m}
\eeq
where the state vector $x(t) \in R^n$ is available for measurement and
is used for generating a state feedback control signal $u(t) \in R^M$,
and $A \in R^{n\times n}$ and $B \in R^{n \times M}$ are unknown
parameter matrices. 

The control objective is to design $u(t)$ to ensure closed-loop system
signal boundedness and asymptotic $x(t)$ tracking the state vector
$x_{m}(t) \in R^n$ of a reference model system
\beq
\left.
\begin{array}{c}
\dot{x}_m(t) \\
x_m(t+1)
\end{array}
\right\} 
= A_{m} x_{m}(t) + B_{m} r(t),
\label{42.3m}
\eeq
where $A_{m} \in R^{n \times n}$ is a constant and stable matrix,
$B_{m} \in R^{n \times M}$ is a constant matrix, and $r(t) \in R^M$ is
a bounded reference input for a desired $x_m(t)$.

\medskip
The state feedback control law structure is 
\beq
u(t) = K_1^{T}(t) x(t) + K_{2}(t) r(t),
\label{4sfcm}
\eeq
where $K_{1}(t) \in R^{n \times M}$ and $K_{2}(t) \in R^{M \times M}$
are estimates of some nominal parameter matrices $K_1^*$ and $K_2^*$
to be defined in the following assumptions (similar to Assumptions (A1)-(A2)):

\begin{description}
\item[] {\bf Assumption (A1M)}: There exist a constant matrix $K_{1}^{*} \in R^{n \times M}$ and a
 nonsingular constant matrix $K_{2}^{*} \in R^{M \times M}$ such that
\beq
A + B K_{1}^{*T} = A_{m},\;
B K_{2}^{*} = B_{m}. 
\label{42.5m}
\eeq
\item[] {\bf Assumption (A2M)}: In Assumption (A1M), $K_2^* =
  \diag\{k_{21}^*, k_{22}^*, \ldots, k_{2M}^*\}$, and
  $\sign[k_{2i}^*]$, $i=1,2,\ldots, M$, are known.
\end{description}

For a continuous-time direct adaptive control design, the following
condition is used \cite{t03}:

\begin{description}
\item[] {\bf Assumption (A2Mc)}: A matrix $S_p \in R^{M \times M}$ is known such that 
$M_{s} = K_{2}^{*} S_p = M_s^T > 0$. 
\end{description}

Assumption (A2M) or (A2Mc) is a generalization of Assumption (A2) to the
multi-input case: if $M = 1$, then $K_2^* = k_2^* \in R$ and $S_p =
\sign[k_2^*]$. While $S_p$ in Assumption (A2Mc) may be complicated to
specify for a general and unknown matrix $K_2^*$, if $K_2^*$ meets
Assumption (A2M): $K_2^* = 
\diag\{k_{21}^*, k_{22}^*, \ldots, k_{2M}^*\}$, then $S_p$ is simple,
for example,
\beq
S_p = \diag\{\sign[k_{21}^*], \sign[k_{22}^*], \ldots,
\sign[k_{2M}^*]\}.
\eeq

\subsection{Designs for Continuous-Time Systems}
\label{4Designs for Continuous-Time SystemsMIMO}
For a continuous-time multi-input plant: $\dot{x}(t) = A x(t) + B
u(t)$, there are two designs: a direct adaptive
control design \cite{t03} and an indirect adaptive control design
\cite{na89}. Next, we present such designs with revisions, to solve 
some additional relevant parameter projection issues.

\subsubsection{Direct Adaptive Control Design}
For the state tracking error $e(t) = x(t) - x_m(t)$, using
(\ref{42.2m})-(\ref{42.5m}) including the control law (\ref{4sfcm}),
in the continuous-time case, we can derive the tracking error equation
\beq
\dot{e}(t) = A_{m} e(t) + B_{m} \left(K_{2}^{*-1} 
\tilde{K}_{1}^{T}(t) x(t) + K_{2}^{*-1} \tilde{K}_{2}(t) r(t)\right),
\label{42.6m}
\eeq
where 
$\tilde{K}_{1}(t) = K_{1}(t) - K_{1}^{*}$ and  
$\tilde{K}_{2}(t) = K_{2}(t) - K_{2}^{*}$.

\medskip
We choose the adaptive laws for the estimates $K_1(t)$ and $K_2(t)$ as
\bea
\label{4dk1m}
\dot{K}_{1}^{T}(t) \ts = \ts - S_p^{T} B_{m}^{T} P e(t) x^{T}(t)
\\*[0.1in]
\dot{K}_{2}(t) \ts = \ts - S_p^{T} B_{m}^{T} P e(t) r^{T}(t),
\label{4dk2m}
\eea
where $P = P^{T} > 0$ satisfying $P A_{m} + A_{m}^{T} P = - Q$ for a
chosen $Q = Q^{T} > 0$, and $S_p$ satisfies the condition in Assumption
(A2Mc).

\slp
\slp
The time-derivative of the positive definite function
\beq
V = e^{T} P e + \tr[\tilde{K}_{1} M_{s}^{-1} \tilde{K}_{1}^{T}] + 
\tr[\tilde{K}_{2}^{T} M_{s}^{-1} \tilde{K}_{2}], 
\eeq
can be derived as $\dot{V} = - e^{T}(t) Q e(t) \leq 0$, from which 
we have that $e(t)$, $\tilde{K}_1(t)$ and $\tilde{K}_2(t)$ are
bounded and $e(t) \in L^2$, that is, $x(t)$, $K_1(t)$ and
$K_2(t)$ are bounded, and so is $u(t)$, that is, all closed-loop
signals are bounded. From (\ref{42.6m}), it follows that $\dot{e}(t)$ is
bounded (so that $e(t)$ is uniformly continuous), and with
$e(t) \in L^2$ and from Barbalat lemma \cite{is96}, that 
$\lim_{t \rightarrow \infty} e(t) = 0$.

\begin{rem}
\rm
If Assumption (A2M) is used, that is, when $K_2^* = \diag\{k_{21}^*,
k_{22}^*, \ldots, k_{2M}^*\}$, then 
\beq
S_p = \diag\{\sign[k_{21}^*] \gamma_1, \sign[k_{22}^*] \gamma_2, \ldots,
\sign[k_{2M}^* \gamma_M]\}
\eeq
with $\gamma_i > 0$, $i=1,2,\ldots, M$, is a choice of $S_p$ for the
adaptive laws (\ref{4dk1m})-(\ref{4dk2m}) which can remain in their
forms or have $K_2(t)$ be projected to be diagonal: all non-diagonal
elements of $K_2(t)$ are set to be zero directly. This follows from
the sepcial parameter projection setting: the initial values and lower
and upper bounds of those non-diagonal elements of $K_2(t)$ are set
to be zero and the derivatives of those elements are made zero by
the corresponding projection signals. \hspace*{\fill} $\fbox{}$
\end{rem}

\subsubsection{Indirect Adaptive Control Design}
An indirect adaptive control design consists of several steps.

\bigskip
{\bf Plant parametrization}. From Assumption (A1M), we express
\beq
A = A_m - B_m \Theta_1^{*T},\;B = B_m \Theta_2^*,
\label{4AnBm}
\eeq
with $\Theta_1^* = K_1^* (K_2^{*-1})^T,\;\Theta_2^* = K_2^{*-1}$, and 
parametrize the plant (\ref{42.2m}) as
\bea
\dot{x}(t) \ts = \ts A x(t) + B u(t) = A_m x(t) + B_m (\Theta_2^{*} u(t) -
\Theta_1^{*T} x(t)).
\label{4param}
\eea

\medskip
{\bf Parameter estimation}. Letting $\Theta_1(t)$ and $\Theta_2(t)$ be
the estimates of the unknown $\Theta_1^*$ and $\Theta_2^*$, 
we first design an adaptive {\it a posteriori} state estimator for $x(t)$:
\beq 
\dot{\hat{x}}(t) = A_m \hat{x}(t) + B_m
(\Theta_2(t) u(t) - \Theta_1^{T}(t) x(t)). 
\label{4estm1} 
\eeq 
For the state estimation error $e_x(t) = \hat{x}(t) - x(t)$,
we obtain the error equation
\beq \dot{e}_x(t) = A_m e_x(t) + B_m ((\Theta_2(t) -
\Theta_2^*) u(t) - (\Theta_1(t) - \Theta_1^*)^T x(t)). \label{4eeem1}
\eeq

We then choose the adaptive laws for $\Theta_1(t)$ and $\Theta_2(t)$:
\bea
\dot{\Theta}_1(t) \ts  = \ts \Gamma_{1} x(t) e_x^T(t)P B_m \label{4theta1dm}\\*[0.05in]
\dot{\Theta}_2(t) \ts = \ts - \Gamma_{2} B_m^T P e_x(t) u^T(t) + F_2(t), \label{4theta2dm}
\eea
where $\Gamma_{1} = \Gamma_{1}^{T} > 0$, $\Gamma_{2} > 0$ is diagonal,
$P = P^{T} > 0$ satisfying $P A_{m} + A_{m}^{T} P = - Q$ for a chosen
$Q = Q^{T} > 0$, and $F_2(t)$ is a projection signal to be designed.

\medskip
For the positive definite function
\beq
V = e_x^T P e_x + \tr[(\Theta_1 - \Theta_1^*)^T\Gamma_1^{-1}(\Theta_1
- \Theta_1^*)] + \tr[(\Theta_2 - \Theta_2^*)^T\Gamma_2^{-1} (\Theta_2
- \Theta_2^*),
\eeq
we derive its time-derivative as
$\dot{V} = - e_x^T Q e_x \leq 0$, from which we conclude that $\Theta_1(t)$,
$\Theta_2(t)$ and $e_x(t)$ are all bounded, and that $e_x(t) \in L^2$.

\begin{rem}
\rm
The adaptive law for $\Theta_1(t)$ may also be chosen as
\beq
\dot{\Theta}_1^T(t)  = \Gamma_{1} B_m^T P e_x(t) x^T(t) \label{4theta1dm1}
\eeq
where $\Gamma_{1} = \Gamma_{1}^{T} > 0$ has different dimensions from
that in (\ref{4theta1dm}).

\medskip
In this case, we consider the positive definite function
\beq
V = e_x^T P e_x + \tr[(\Theta_1 - \Theta_1^*)\Gamma_1^{-1}(\Theta_1
- \Theta_1^*)^T] + \tr[(\Theta_2 - \Theta_2^*)^T\Gamma_2^{-1} (\Theta_2
- \Theta_2^*),
\eeq
and can also derive its time-derivative as
$\dot{V} = - e_x^T Q e_x \leq 0$. \hspace*{\fill} $\fbox{}$
\end{rem}

{\bf Control law}. The adaptive control law has the form (\ref{4sfcm}):
\beq
u(t) = K_1^T(t) x(t) + K_2(t) r(t),
\label{4aclm}
\eeq
where, for an indirect design, its parameters are calculated from
\beq
K_1^T(t) = \Theta_2^{-1}(t)
\Theta_1^T(t),\;K_2(t) = \Theta_2^{-1}(t).
\eeq
To implement this control law, the parameter estimate $\Theta_2(t)$
needs to be ensured to be nonsingular for all $t \geq 0$, 
by using parameter projection on $\Theta_2(t)$. While parameter
projection can be easily done if $K_2^*$ is diagonal or triangular
(and so is $\Theta_2(t)$, as $\Theta_2^* = K_2^{*-1}$), it may also be
done using some relevant knowledge of a more general matrix $K_2^*$. 

\medskip
With the control law (\ref{4aclm}), 
the estimator equation (\ref{4estm1}) becomes
\beq
\dot{\hat{x}}(t) = A_m \hat{x}(t) + B_m r(t),
\label{4est1m}
\eeq
that is, $\hat{x}(t)$ is bounded so that $x(t) = 
\hat{x}(t) - e_x(t)$, $u(t)$ in (\ref{4aclm}) and $\dot{x}(t)$ in (\ref{42.2m}) are bounded, 
and $\lim_{t \rightarrow \infty} (\hat{x}(t) - x_m(t)) = 0$
exponentially so that $\hat{x}(t) - x_m(t) \in L^2$. Hence we have
that $x(t) - x_m(t) \in L^2$, and, with $\dot{x}(t) - \dot{x}_m(t) \in
L^\infty$, that $\lim_{t \rightarrow \infty} (x(t) - x_m(t)) = 0$.

\bigskip
{\bf Parameter projection under Assumption (A2M)}. For $K_2^* =
\diag\{k_{21}^*, k_{22}^*, \ldots, k_{2M}^*\}$ with $\sign[k_{2i}^*]$
known, $i=1,2,\ldots, M$, we also assume:

\begin{description}
\item[] {\bf Assumption (A3M)}: Upper bounds $k_{2i}^b$ of
$|k_{2i}^*|$: $k_{2i}^b \geq |k_{2i}^*|$, $i=1,2,\ldots, M$, are known.
\end{description}

In view of (\ref{4AnBm}), in terms of $\Theta_2^* = K_2^{*-1} = \diag\{\theta_{21}^*,
\theta_{22}^*, \ldots, \theta_{2M}^*\}$ with $\theta_{2i}^* =
1/k_{2i}^*$, Assumption (A2M) implies that $\sign[\theta_{2i}^*]$,
  $i=1,2,\ldots, M$, are known, and Assumption (A3M) implies that
  lower bounds $\theta_{2i}^a = 1/k_{2i}^b$ of $|\theta_{2i}^*|$,
  $i=1,2,\ldots, M$, are known.

For parameter projection under the condition that $K_2^*$ is diagonal
(and so is $\Theta_2^*$, so that $\Theta_2(t)$ should be made to be
diagonal: $\Theta_2(t) = \diag\{\theta_{21}(t), \theta_{22}(t),
\ldots, \theta_{2M}(t)\}$), we set the initial values and the derivatives of the
non-diagonal elements of $\Theta_2(t)$ to be zero, choose $F_2(t)$ in
(\ref{4theta2dm}) to be diagonal: $F_2(t) = \diag\{f_{21}(t),
f_{22}(t), \ldots, f_{2M}(t)\}$, let $G_2(t) =  - \Gamma_{2} B_m^T P e_x(t) u^T(t)$
(with $\Gamma_2 = \Gamma_2 > 0$ being diagonal) in (\ref{4theta2dm})
and denote the diagonal elements of $G_2(t)$ as $g_{2i}(t)$ for
$i=1,2,\ldots, M$, choose $\theta_{2i}(0)$ to be such that
$\sign[\theta_{2i}^*] \theta_{2i}(0) \geq \theta_{2i}^a > 0$, and set 
\beq
f_{2i}(t) = \left\{ \begin{array}{ll}
0  & \mbox{if $\sign[\theta_{2i}^*]\theta_{2i}(t) > \theta_{2i}^a$, or}\\
 & \mbox{if $\sign[\theta_{2i}^*] \theta_{2i}(t) = \theta_{2i}^a$ and
 $\sign[\theta_{2i}^*] g_{2i}(t) \geq 0$}\\*[0.05in]
- g_{2i}(t) & \mbox{otherwise,}
\end{array}
\right.
\label{4f2i(t)}
\eeq
which ensures that $\sign[\theta_{2i}(t)] =
\sign[\theta_{2i}^*]$, $|\theta_{2i}(t)| \geq \theta_{2i}^a > 0$ and 
$(\theta_{2i}(t) - \theta_{2i}^*) f_{2i}(t) \leq 0$.

\subsection{Designs for Discrete-Time Systems}
We now consider the discrete-time version of the plant (\ref{42.2m}):
\beq
x(t+1) = A x(t) + B u(t),
\label{42.2md}
\eeq
with $x(t) \in R^n$ and $u(t) \in R^M$ for $M > 1$, the reference
system (\ref{42.3m}): 
\beq
x_{m}(t+1) = A_{m} x_{m}(t) + B_{m} r(t),
\label{42.3md}
\eeq
with $A_{m} \in R^{n \times n}$ stable, $B_{m} \in R^{n \times M}$,
and the control law (\ref{4sfcm}):
\beq
u(t) = K_1^{T}(t) x(t) + K_{2}(t) r(t),
\label{4sfcmd}
\eeq
with $K_{1}(t) \in R^{n \times M}$ and $K_{2}(t) \in R^{M \times M}$
as the estimates of some nominal parameter matrices $K_1^*$ and
$K_2^*$ satisfying Assumption (A1M): $A + B K_{1}^{*T} = A_{m},\;
B K_{2}^{*} = B_{m}$.

\medskip
The control law (\ref{4sfcmd}), applied to the plant (\ref{42.2md}),
results in
\beq
e(t+1) = A_m e(t) + B_m K_2^{*-1} \tilde{\Theta}^T(t) \omega(t),
\label{4erroreqm}
\eeq
where $\omega(t) = [x^T(t), r^T(t)]^T$ and $\tilde{\Theta}(t)
= \Theta(t) - \Theta^*$ with
\beq
\Theta(t) = [K_1^T(t), K_2(t)]^T,\;
\Theta^* = [K_1^{*T}, K_2^*]^T.
\eeq

The error equation (\ref{4erroreqm}) is based on a direct adaptive
control formulation in which the parameter matrix $\Theta(t)$ to be
updated directly contains the controller parameters $K_1(t)$ and
$K_2(t)$. An indirect adaptive control formulation (see
Section \ref{4Indirect Adaptive Control Design}) has a similar
equation (see (\ref{432eee}) for the case of $M=1$, with $e(t)$
replaced by an estimation error $e_x(t)$, and without the term
$K_2^{*-1}$). Next, we study the design of a gradient algorithm for
discrete-time adaptive control based on such an error equation (to
which a Lyapunov algorithm is not applicable).

\subsubsection{An Illustrative Example}
We consider an example of the error equation (\ref{4erroreqm}):
\beq
e(t+1) = A_m e(t) + B_m K^{*} \tilde{\Theta}^T(t) \omega(t),
\label{4erroreqme}
\eeq
for the case of $n=3$ and $M = 2$: $e(t) \in R^3$, $A_m \in R^{3 \times 3}$, $B_m \in
R^{3 \times 2}$, $K^* \in R^{2 \times 2}$, $\tilde{\Theta}^T(t) 
= [\tilde{\theta}_1(t), \tilde{\theta}_2(t)]^T = (\Theta(t) - \Theta^{*})^T \in
R^{2 \times n_\theta}$, and $\omega(t) \in R^{n_\theta}$. 

\medskip
For $K^* = [k_{ij}^*]$ with $i=1,2, j=1,2$, and
$W_m(z) = (zI - A_m)^{-1} B_m = [w_{ij}(z)]$ with $i=1,2,3$, $j=1,2$, we have the expression
\bea
e(t) \ts = \ts W_m(z)[K^* \tilde{\Theta}^T \omega](t) \nn\\
\ts = \ts \left[ \begin{array}{cc}
w_{11}(z) & w_{12}(z) \\
w_{21}(z) & w_{22}(z) \\
w_{31}(z) & w_{32}(z) 
\end{array}
\right][
\left[ \begin{array}{cc}
k_{11}^* & k_{12}^*\\
k_{21}^* & k_{22}^*\\
\end{array}
\right] \left[ \begin{array}{c}
\tilde{\theta}_1^T \omega \\
\tilde{\theta}_2^T \omega 
\end{array}
\right]](t) \nn\\
\ts = \ts \left[ \begin{array}{cc}
w_{11}(z) k_{11}^* + w_{12}(z) k_{21}^* & w_{11}(z) k_{12}^* + w_{12}(z) k_{22}^* \\
w_{21}(z) k_{11}^* + w_{22}(z) k_{21}^* & w_{21}(z) k_{12}^* + w_{22}(z) k_{22}^* \\
w_{31}(z) k_{11}^* + w_{32}(z) k_{21}^* & w_{31}(z) k_{12}^* + w_{32}(z) k_{22}^*
\end{array}
\right] \left[ \begin{array}{c}
\tilde{\theta}_1^T \omega \\
\tilde{\theta}_2^T \omega 
\end{array}
\right]](t) \nn\\
\ts = \ts \left[ \begin{array}{cc}
(w_{11}(z) k_{11}^* + w_{12}(z) k_{21}^*)[\tilde{\theta}_1^T \omega](t) + 
(w_{11}(z) k_{12}^* + w_{12}(z) k_{22}^*)[\tilde{\theta}_2^T \omega](t) \\ 
(w_{21}(z) k_{11}^* + w_{22}(z) k_{21}^*)[\tilde{\theta}_1^T \omega](t) + 
(w_{21}(z) k_{12}^* + w_{22}(z) k_{22}^*)[\tilde{\theta}_2^T \omega](t) \\
(w_{31}(z) k_{11}^* + w_{32}(z) k_{21}^*)[\tilde{\theta}_1^T \omega](t) +
(w_{31}(z) k_{12}^* + w_{32}(z) k_{22}^*)[\tilde{\theta}_2^T \omega](t) 
\end{array}
\right] \nn\\
\ts \ts \mbox{(it may not lead to a solution for a general $K^* \in R^{2
\times 2}$)}.
\eea
To see this, we examine, for example, 
\beq
(w_{11}(z) k_{11}^* + w_{12}(z) k_{21}^*)[\tilde{\theta}_1^T
  \omega](t) = 
 k_{11}^* w_{11}(z)[\tilde{\theta}_1^T \omega](t) + 
 k_{21}^* w_{12}(z)[\tilde{\theta}_1^T \omega](t).
\eeq
Such a combined signal involves the combined uncertainty of $k_{11}^*$
and $k_{21}^*$, whose sign is uncertain, while the sign of $k_{11}^*$
can be specified from the single parameter $k_{11}^*$.

\medskip
Hence, we need to consider a diagonal $K^* = \diag\{k_{11}^*,
k_{22}^*\}$ and then obtain
\bea
e(t) \ts = \ts W_m(z)[K^* \tilde{\Theta}^T \omega](t) \nn\\
\ts = \ts \left[ \begin{array}{cc}
w_{11}(z) k_{11}^*[\tilde{\theta}_1^T \omega](t) + 
w_{12}(z) k_{22}^*[\tilde{\theta}_2^T \omega](t) \\ 
w_{21}(z) k_{11}^*[\tilde{\theta}_1^T \omega](t) + 
w_{22}(z) k_{22}^*[\tilde{\theta}_2^T \omega](t) \\
w_{31}(z) k_{11}^*[\tilde{\theta}_1^T \omega](t) +
w_{32}(z) k_{22}^*[\tilde{\theta}_2^T \omega](t) 
\end{array}
\right].
\eea
With $e(t) = [e_1(t), e_2(t), e_3(t)]^T$, for $i=1,2,3$, we have 
\bea
e_i(t) \ts = \ts w_{i1}(z) k_{11}^*[\tilde{\theta}_1^T \omega](t) + 
w_{i2}(z) k_{22}^*[\tilde{\theta}_2^T \omega](t) \nn\\
 \ts = \ts k_{11}^* w_{i1}(z)[\tilde{\theta}_1^T \omega](t) + 
k_{22}^* w_{i2}(z)[\tilde{\theta}_2^T \omega](t).
\label{4miei(t)}
\eea

For $i=1,2,3$, we introduce the estimation errors
\beq
\epsilon_i(t) = e_i(t) + k_{11}(t) \xi_{i1}(t) + k_{22}(t) \xi_{i2}(t), 
\label{4miepsi(t)}
\eeq
where $k_{11}(t)$ and $k_{22}(t)$ are the estimates of $k_{11}^*$ and
$k_{22}^*$, and
\bea
\label{4mixii1}
\xi_{i1}(t) \ts = \ts \theta_1^T(t) \zeta_{i1}(t) -
w_{i1}(z)[\theta_1^T \omega](t),\;\zeta_{i1}(t) =
w_{i1}(z)[\omega](t) \\
\xi_{i2}(t) \ts = \ts \theta_2^T(t) \zeta_{i2}(t) -
w_{i2}(z)[\theta_2^T \omega](t),\;\zeta_{i2}(t) =
w_{i2}(z)[\omega](t).
\label{4mixii2}
\eea
From (\ref{4miei(t)})-(\ref{4mixii2}), we derive
\bea
\epsilon_i(t) \ts = \ts  k_{11}^* (\theta_1(t) - \theta_1^*)^T \zeta_{i1}(t) +
(k_{11}(t) - k_{11}^*) \xi_{i1}(t) \nn\\
\ts \ts + 
k_{22}^* (\theta_2(t) - \theta_2^*)^T \zeta_{i2}(t) +
(k_{22}(t) - k_{22}^*) \xi_{i2}(t).
\label{4miepsi(t)e}
\eea
 
Consider the cost function
\beq
J(\theta_1, \theta_2) = \frac{\sum_{i=1}^3 \epsilon_i^2}{m^2}
\eeq
and obtain its gradients
\beq
\frac{\partial J}{\partial \theta_1} = \frac{k_{11}^* \sum_{i=1}^3 \epsilon_i \zeta_{i1}}{m^2(t)},\;
\frac{\partial J}{\partial \theta_2} = \frac{k_{22}^* \sum_{i=1}^3 \epsilon_i \zeta_{i2}}{m^2(t)}
\eeq
\beq
\frac{\partial J}{\partial k_{11}} = \frac{\sum_{i=1}^3 \epsilon_i \xi_{i1}}{m^2(t)},\;
\frac{\partial J}{\partial k_{22}} = \frac{\sum_{i=1}^3 \epsilon_i \xi_{i2}}{m^2(t)}.
\eeq
This motivates us to choose the adaptive laws
\bea
\label{4theta1t+1}
\theta_1(t+1) \ts = \ts \theta_1(t)
- \frac{\sign[k_{11}^*] \Gamma_1 \sum_{i=1}^3 \epsilon_i \zeta_{i1}}{m^2(t)}
\\*[0.05in]
\theta_2(t+1) \ts = \ts \theta_2(t)
- \frac{\sign[k_{22}^*] \Gamma_2 \sum_{i=1}^3 \epsilon_i \zeta_{i2}}{m^2(t)}
\\*[0.05in]
k_{11}(t+1) \ts = \ts k_{11}(t)
- \frac{\gamma_1 \sum_{i=1}^3 \epsilon_i \xi_{i1}}{m^2(t)}
\\*[0.05in]
k_{22}(t+1) \ts = \ts k_{22}(t)
- \frac{\gamma_2 \sum_{i=1}^3 \epsilon_i \xi_{i2}}{m^2(t)},
\label{4k22t+1}
\eea
where $0 < \Gamma_i = \Gamma_i^T < 2/|k_{ii}^*|$ and $0< \gamma_i < 2$,
 $i=1,2$, and 
\beq
m(t) = \sqrt{1 + \sum_{i=1}^3 \left(\zeta_{i1}^T(t) \zeta_{i1}(t) + 
\zeta_{i2}^T(t) \zeta_{i2}(t) + \xi_{i1}^2(t) + \xi_{i2}^2(t)\right)}.
\eeq

Consider the positive definite function
\beq
V(\tilde{\theta}_1, \tilde{\theta}_2, \tilde{k}_{11}, \tilde{k}_{22})
 = |k_{11}^*|\tilde{\theta}_1^{T} \Gamma_1^{-1} \tilde{\theta}_1 
+ |k_{22}^*|\tilde{\theta}_2^{T} \Gamma_2^{-1} \tilde{\theta}_2 
+ \gamma_1^{-1} \tilde{k}_{11}^{2} + \gamma_2^{-1} \tilde{k}_{22}^{2} 
\label{432A.40e}
\eeq
where the parameter errors are
\beq
\tilde{\theta}_i(t) = \theta_i(t) - \theta_i^{*},\;
\tilde{k}_{ii}(t) = k_{ii}(t) - k_{ii}^{*},\;i=1,2.
\label{432A.41e}
\eeq
The time-increment of $V$ can be derived as
\bea
\ts \ts V(\tilde{\theta}_1(t+1), \tilde{\theta}_2(t+1), \tilde{k}_{11}(t+1), \tilde{k}_{22}(t+1))
-
  V(\tilde{\theta}_1(t), \tilde{\theta}_2(t), \tilde{k}_{11}(t), \tilde{k}_{22}(t)) \nn\\
\ts \leq \ts 
 - (2 - \gamma_0)
\frac{\sum_{i=1}^3 \epsilon_i^2(t)}{m^2(t)} \leq 0,
\eea
for some $\gamma_0 \in (0, 2)$, which leads to the desired properties:

\begin{lem}
\label{4lemmadie}
The adaptive laws (\ref{4theta1t+1})-(\ref{4k22t+1}) ensure:

\medskip
(i) $\theta_i(t)$ and $k_{ii}(t)$, $i=1,2$, and $\frac{\sum_{i=1}^3 \epsilon_i^2(t)}{m^2(t)}$ are 
 bounded; and 

\medskip
(ii) $\frac{\sum_{i=1}^3 \epsilon_i^2(t)}{m^2(t)} \in L^1$, 
$\frac{\epsilon_i(t)}{m(t)} \in L^2$, $\theta_i(t+1)
- \theta_i(t) \in L^2$, $k_{ii}(t+1) - k_{ii}(t) \in L^{2}$, $i=1,2$.
\end{lem}

\medskip
{\bf Summary}. For an error equation (\ref{4erroreqme}) with a matrix
$K^*$ of the form:
\beq
e(t+1) = A_m e(t) + B_m K^{*} \tilde{\Theta}^T(t) \omega(t),
\eeq
a gradient adaptive law design requires $K^*$ to be diagonal. This
applies to a direct adaptive control design for a multi-input
discrete-time system: the nominal parameter matrix $K_2^*$ needs 
to be diagonal: $K_2^* = \diag\{k_{21}^*, k_{22}^*, \ldots,
k_{2M}^*\}$ (see Assumption (A2M)). 

For an indirect adaptive control design, such an equation has $K^* =
I$, and the input-output model of the above equation  is
\beq
e(t) =  W_m(z)[\tilde{\Theta}^T \omega](t),\;W_m(z) = (zI - A_m)^{-1}
 B_m,
\eeq
and the adaptive scheme for estimating $\Theta(t)$ can be designed
following the technique developed in \cite{tl99}. However, for
indirect adaptive control, a submatrix $\Theta_2$ (also corresponding
to $K_2^*$) of $\Theta$ needs to be made to be nonsingular
by parameter projection (similar to that in Section \ref{4Indirect Adaptive Control
Design}, where $\theta_2(t)$ needs to be nonzero, for the case of $M =
1$), which can be easily done if $K_2^*$ is
diagonal.

\subsubsection{Direct Adaptive Control Design}
We follow the tracking error equation (\ref{4erroreqm}):
\beq
e(t+1) = A_m e(t) + B_m K_2^{*-1} \tilde{\Theta}^T(t) \omega(t),
\label{4erroreqm1}
\eeq
and, based on the above analysis, in addition to Assumption (A2M),
also assume:

\begin{description}
\item[] {\bf Assumption (A4M)}: 
Lower bounds $k_{2i}^a > 0$ of $|k_{2i}^*|$: $|k_{2i}^*| \geq
k_{2i}^a$, $i=1,2,\ldots, M$, are known.
\end{description}

Then, denoting $\rho_i^* = 1/k_{2i}^*$, $i=1,2,\ldots, M$, and 
\beq
\Theta(t) = [\theta_1(t), \theta_2(t), \ldots, \theta_M(t)],\;
\Theta^* = [\theta_1^*, \theta_2^*, \ldots, \theta_M^*],
\eeq
with $W_m(z) = (zI - A_m)^{-1} B_m$,  
$\tilde{\theta}_i(t) = \theta_i(t) - \theta_i^*$, $i=1,2,\ldots,
M$, and $K_2^{*-1} = \diag\{\rho_1^*, \rho_2^*, \ldots, \rho_M^*\}$, 
we express (\ref{4erroreqm1}) in the input-output form as
\beq
e(t) = W_m(z)[\left[ \begin{array}{c}
\rho_1^* \tilde{\theta}_1^T \omega\\
\rho_2^* \tilde{\theta}_2^T \omega\\
\vdots \\
\rho_M^* \tilde{\theta}_M^T \omega
\end{array}
\right]](t),
\label{4e(t)Wm(z)}
\eeq
which, with $e(t) = [e_1(t), \ldots, e_n(t)]^T$ and $W_m(z) =
[w_{ij}(z)]$, $i=1,2,\ldots, n$, $j = 1, 2, \ldots, M$, can be further
written as
\bea 
e_i(t) \ts = \ts w_{i1}(z)[\rho_1^* \tilde{\theta}_1^T \omega](t) + 
w_{i2}(z)[\rho_2^* \tilde{\theta}_2^T \omega](t) + \cdots + 
w_{iM}(z)[\rho_M^* \tilde{\theta}_M^T \omega](t),\;i=1,2,\ldots, n\nn\\
 \ts = \ts \rho_1^* w_{i1}(z)[\tilde{\theta}_1^T \omega](t) + 
\rho_2^* w_{i2}(z)[\tilde{\theta}_2^T \omega](t) + \cdots + 
\rho_M^* w_{iM}(z)[\tilde{\theta}_M^T \omega](t).
\label{4miei(t)m}
\eea

We introduce the estimation errors
\beq
\epsilon_i(t) = e_i(t) + \rho_{1}(t) \xi_{i1}(t)
+ \rho_{2}(t) \xi_{i2}(t) + \cdots + \rho_{M}(t) \xi_{iM}(t),
\label{4miepsi(t)m}
\eeq
where $\rho_{j}(t)$, $j=1,2,\ldots, M$, are the estimates of
$\rho_{j}^*$, and
\beq
\xi_{ij}(t) = \theta_j^T(t) \zeta_{ij}(t) -
w_{ij}(z)[\theta_j^T \omega](t),\;\zeta_{ij}(t) =
w_{ij}(z)[\omega](t). 
\label{4mixii1m}
\eeq
From (\ref{4miei(t)m})-(\ref{4mixii1m}), we derive
\bea 
\epsilon_i(t) \ts = \ts  \rho_1^* (\theta_1(t) - \theta_1^*)^T \zeta_{i1}(t) +
(\rho_{1}(t) - \rho_{1}^*) \xi_{i1}(t) + \cdots \nn\\
\ts \ts + 
\rho_{M}^* (\theta_M(t) - \theta_M^*)^T \zeta_{iM}(t) +
(\rho_{M}(t) - \rho_{M}^*) \xi_{iM}(t).
\label{4miepsi(t)em}
\eea
 
We choose the adaptive laws
\bea
\label{4theta1t+1m}
\theta_i(t+1) \ts = \ts \theta_i(t)
- \frac{\sign[\rho_{i}^*] \Gamma_i \sum_{k=1}^n \epsilon_k \zeta_{ki}}{m^2(t)}
\\*[0.05in]
\rho_{i}(t+1) \ts = \ts \rho_i(t)
- \frac{\gamma_i \sum_{k=1}^n \epsilon_k \xi_{ki}}{m^2(t)},
\label{4k22t+1m}
\eea
where $0 < \Gamma_i = \Gamma_i^T |\rho_i^*| < 2 I$ and $0< \gamma_i < 2$,
 $i=1,2, \ldots, M$, and 
\beq
m(t) = \sqrt{1 + \sum_{i=1}^n \sum_{j=1}^M
(\zeta_{ij}^T(t) \zeta_{ij}(t) + \xi_{ij}^2(t))}. 
\eeq
In view of the definition of $\rho_i^* = 1/k_{2i}^*$ and Assumption (A4M), we
can choose $0 < \Gamma_i = \Gamma_i^T < k_{2i}^a I$ with $|k_{2i}^*|
\geq k_{2i}^a > 0$, $i=1,2,\ldots, M$, for $k_{2i}^a$ known.

\medskip
This adaptive scheme has the same properties as that in
Lemma \ref{4lemmadie}:

\medskip
\medskip
(i) $\theta_i(t)$ and $\rho_i(t)$, $i=1,2, \ldots, M$, and
$\frac{\sum_{k=1}^n \epsilon_k^2(t)}{m^2(t)}$ are bounded; and 

\medskip
(ii) $\frac{\sum_{k=1}^n \epsilon_k^2(t)}{m^2(t)} \in L^1$, 
$\frac{\epsilon_i(t)}{m(t)} \in L^2$, $\theta_i(t+1)
- \theta_i(t) \in L^2$, and $\rho_i(t+1) - \rho_{i}(t) \in L^{2}$,
$i=1,2, \ldots, M$.

\begin{rem}
\rm
Under Assumption (A2M), $K_2^* = \diag\{k_{21}^*, k_{22}^*, \ldots,
k_{2M}^*\}$, the matrix $K_2(t)$ in $\Theta(t) = [K_1^T(t), K_2(t)]^T$
for (\ref{4erroreqm1}) can also be made to be diagonal by parameter
projection. 

For parameter projection, the gain matrix $\Gamma_i$ in
(\ref{4theta1t+1m}) is chosen as $\Gamma_i = \diag\{\Gamma_{i1},
\Gamma_{i2}\}$ with $\Gamma_{i1} \in R^{n \times n}$ and $\Gamma_{i2}
\in R^{M \times M}$ which is diagonal. Then, with 
\beq
\Theta(t) = [\theta_1(t), \theta_2(t), \ldots, \theta_M(t)] = 
\left[ \begin{array}{c}
K_1(t)\\
K_2^T(t)
\end{array}
\right],
\eeq
we can set the non-diagonal elements of $K_2(t)$ as zero. \hspace*{\fill} $\fbox{}$
\end{rem}

\subsubsection{Indirect Adaptive Control Design}
From Assumption (A1M), we express
\beq
A = A_m - B_m \Theta_1^{*T},\;B = B_m \Theta_2^*,
\label{4AnBmd}
\eeq
with $\Theta_1^* = K_1^* (K_2^{*-1})^T,\;\Theta_2^* = K_2^{*-1}$, and 
parametrize the plant (\ref{42.2md}) as
\bea
x(t+1) \ts = \ts A x(t) + B u(t) = A_m x(t) + B_m (\Theta_2^{*} u(t) -
\Theta_1^{*T} x(t)).
\label{4paramd}
\eea

\medskip
{\bf Parameter estimation}. Letting $\Theta_1(t)$ and $\Theta_2(t)$ be
the estimates of the unknown parameters $\Theta_1^*$ and $\Theta_2^*$, 
we design a discrete-time adaptive {\it a posteriori} state estimator for $x(t)$:
\beq 
\hat{x}(t+1) = A_m \hat{x}(t) + B_m
(\Theta_2(t) u(t) - \Theta_1^{T}(t) x(t)). 
\label{4estm1d} 
\eeq 
For the state estimation error $e_x(t) = \hat{x}(t) - x(t)$,
we obtain
\beq 
e_x(t+1) = A_m e_x(t) + B_m ((\Theta_2(t) -
\Theta_2^*) u(t) - (\Theta_1(t) - \Theta_1^*)^T x(t)), \label{4eeem1d}
\eeq
which, with $\omega(t) = [-x^T(t), u^T(t)]^T$ and $\tilde{\Theta}(t)
= \Theta(t) - \Theta^*$ for 
\beq
\Theta(t) = [\Theta_1^T(t), \Theta_2(t)]^T,\;
\Theta^* = [\Theta_1^{*T}, \Theta_2^*]^T,
\eeq
can be expressed in the form of (\ref{4erroreqm}) or (\ref{4erroreqm1}) without $K_2^{*-1}$:
\beq
e_x(t+1) = A_m e_x(t) + B_m \tilde{\Theta}^T(t) \omega(t).
\label{4erroreqmd}
\eeq

With $W_m(z) = (zI - A_m)^{-1} B_m$, and 
$\tilde{\theta}_i(t) = \theta_i(t) - \theta_i^*$, $i=1,2,\ldots, M$, for
\beq
\Theta(t) = [\theta_1(t), \theta_2(t), \ldots, \theta_M(t)],\;
\Theta^* = [\theta_1^*, \theta_2^*, \ldots, \theta_M^*],
\eeq
we express (\ref{4erroreqmd}) as 
\beq
e_x(t) = W_m(z)[\left[ \begin{array}{c}
 \tilde{\theta}_1^T \omega\\
 \tilde{\theta}_2^T \omega\\
\vdots \\
 \tilde{\theta}_M^T \omega
\end{array}
\right]](t),\label{4erroreqmd1}
\eeq

With $e_x(t) = [e_{x1}(t), e_{x2}(t), \ldots, e_{xn}(t)]^T$ and $W_m(z) =
[w_{ij}(z)]$, $i=1,2,\ldots, n$, $j = 1, 2, \ldots, M$,
(\ref{4erroreqmd1}) can be written as
\beq
e_{xi}(t) = w_{i1}[\tilde{\theta}_1^T \omega](t) + 
 w_{i2}[\tilde{\theta}_2^T \omega](t) + \cdots + 
 w_{iM}[\tilde{\theta}_M^T \omega](t).
\label{4miei(t)md}
\eeq

Similarly, we introduce the estimation errors
\beq
\epsilon_i(t) = e_{xi}(t) + \xi_{i1}(t)
+ \xi_{i2}(t) + \cdots + \xi_{iM}(t),
\label{4miepsi(t)md}
\eeq
where
\beq
\xi_{ij}(t) = \theta_j^T(t) \zeta_{ij}(t) -
w_{ij}(z)[\theta_j^T \omega](t),\;\zeta_{ij}(t) =
w_{ij}(z)[\omega](t). 
\label{4mixii1md}
\eeq
From (\ref{4miei(t)md})-(\ref{4mixii1md}), we derive
\beq 
\epsilon_i(t) = (\theta_1(t) - \theta_1^*)^T \zeta_{i1}(t) + \cdots +
 (\theta_M(t) - \theta_M^*)^T \zeta_{iM}(t).
\label{4miepsi(t)emd}
\eeq
 
For $i=1,2,\ldots, M$, we choose the adaptive laws
\beq
\theta_i(t+1) = \theta_i(t)
- \frac{\Gamma_i \sum_{k=1}^n \epsilon_k \zeta_{ki}}{m^2(t)}
\label{4theta1t+1md}
\eeq
where $0 < \Gamma_i = \Gamma_i^T < 2 I$, and
\beq
m(t) = \sqrt{1 + \sum_{i=1}^n \sum_{j=1}^M
(\zeta_{ij}^T(t) \zeta_{ij}(t) + \xi_{ij}^2(t))}. 
\eeq

\medskip
This adaptive scheme has the same properties as that in
Lemma \ref{4lemmadie}:

\medskip
\medskip
(i) $\theta_i(t)$, $i=1,2, \ldots, M$, and
$\frac{\sum_{k=1}^n \epsilon_k^2(t)}{m^2(t)}$ are bounded; and 

\medskip
(ii) $\frac{\sum_{k=1}^n \epsilon_k^2(t)}{m^2(t)} \in L^1$, 
$\frac{\epsilon_i(t)}{m(t)} \in L^2$, and $\theta_i(t+1)
- \theta_i(t) \in L^{2}$, $i=1,2,\ldots, M$.

\bigskip
{\bf Parameter projection}. To use $\Theta(t)$ for control design, we
need to ensure $\Theta_2(t)$ in 
$\Theta(t) = [\Theta_1^T(t), \Theta_2(t)]^T$ is nonsingular. This may
be achieved by using parameter projection which can be easily done if
$K_2^*$ is diagonal or triangular (and so is $\Theta_2$) and if the
signs and the upper bounds $k_{2i}^b \geq |k_{2i}^*|$ of the diagonal
elements $k_{2i}^*$ of $K_2^*$ are known, $i=1,2,\ldots, M$.

\medskip
For parameter projection design, we recall
\beq
\Theta(t) = [\theta_1(t), \theta_2(t), \ldots, \theta_M(t)] =
[\Theta_1^T(t), \Theta_2(t)]^T = \left[ \begin{array}{c}
\Theta_1(t)\\
\Theta_2^T(t)
\end{array}
\right],
\eeq
and modify the adaptive laws (\ref{4theta1t+1md}) as
\beq
\theta_i(t+1) = \theta_i(t)
- \frac{\Gamma_i \sum_{k=1}^n \epsilon_k \zeta_{ki}}{m^2(t)} + f_i(t),
\label{4theta1t+1mdp}
\eeq
where $\Gamma_i = \diag\{\Gamma_{i1}, \Gamma_{i2}\}$ with
$\Gamma_{i1} \in R^{n \times n}$ such that $\Gamma_{i1}
= \Gamma_{i1}^T > 0$ and $\Gamma_{i2} \in R^{M \times M}$ such that
$\Gamma_{i2} > 0$ is diagonal (corresponding to the last $M$
components of $\theta_i(t)$), to design
the projection functions $f_i(t)$ (whose first $n$
components are set to be zero), $i=1,2,\ldots, M$.

\medskip
Under Assumption (A2M), $\Theta_2^* = K_2^{*-1}$ is diagonal, and we
also choose $\Theta_2(t)$ to be diagonal, which can be done by setting
the non-diagonal elements of $\Theta_2(0)$ and $\Theta_2(t)$ to be
zero for all $t > 0$. To design the parameter projection signals $f_i(t)$,
$i=1,2,\ldots, M$, we denote 
\beq
F(t) = [f_1(t), f_2(t), \ldots, f_M(t)] = 
\left[\begin{array}{c}
F_1(t)\\
F_2(t)
\end{array}
\right],
\eeq
where $F_1(t) \in R^{n \times M}$ is set to be zero and $F_2(t) \in
R^{M \times M}$ is a diagonal matrix whose diagonal elements are
denoted as $f_{2i}(t)$, $i=1,2,\ldots, M$. We also denote 
\beq
g_i(t) = - \frac{\Gamma_i \sum_{k=1}^n \epsilon_k \zeta_{ki}}{m^2(t)}
\eeq
and form 
\beq
G(t) = [g_1(t), g_2(t), \ldots, g_M(t)] = 
\left[\begin{array}{c}
G_1(t)\\
G_2(t)
\end{array}
\right],
\eeq
where $G_1(t) \in R^{n \times M}$, and $G_2(t) \in
R^{M \times M}$ with diagonal elements $g_{2i}(t)$, $i=1,2,\ldots, M$. 

We denote the diagonal elements of $\Theta_2(t)$ as $\theta_{2i}(t)$,
$i=1,2,\ldots, M$, and choose $\theta_{2i}(0)$ such that
$\sign[\theta_{2i}(0)] = \sign[\theta_{2i}^*] = 
\sign[k_{2i}^*]$ and $|\theta_{2i}(0)| \geq \theta_{2i}^a$, 
where the lower bounds $\theta_{2i}^a = 1/k_{2i}^b$ of
$|\theta_{2i}^*|$, $i=1,2,\ldots, M$, are known (see Assumption
(A3M)). We then set the projection signals $f_{2i}(t)$ as
\beq
f_{2i}(t) = \left\{ \begin{array}{ll}
0 & \mbox{if $\sign[k_{2i}^*](\theta_{2i}(t)+g_{2i}(t)) \geq \theta_{2i}^a$, }\\
\theta_{2i}^a - \theta_{2i}(t) - g_{2i}(t) & \mbox{otherwise,}
\end{array}
\right.
\label{432projm}
\eeq
to ensure that $\sign[\theta_{2i}(t)] = \sign[\theta_{2i}^*] =
\sign[k_{2i}^*]$ and $|\theta_{2i}(t)| \geq \theta_{2i}^a$, and 
\beq
(\theta_{2i}(t) - \theta_{2i}^* + g_{2i}(t) + f_{2i}(t)) f_{2i}(t) \leq 0.
\label{4125pp1dm}
\eeq

\bigskip
{\bf Control law}. The adaptive control law has the form (\ref{4sfcm}):
\beq
u(t) = K_1^T(t) x(t) + K_2(t) r(t),
\label{4aclmd}
\eeq
where, for this indirect adaptive control design, its parameters are calculated from
\beq
K_1^T(t) = \Theta_2^{-1}(t)
\Theta_1^T(t),\;K_2(t) = \Theta_2^{-1}(t),
\eeq
where the parameter matrix $\Theta_2(t)$ is ensured to be
nonsingular for all $t \geq 0$, by parameter projection.

\subsubsection{Applications to Continuous-Time Systems}
\label{4Application to Continuous-Time Systems}
The developed direct and indirect gradient algorithm based adaptive
control schemes present solutions to the open adaptive state tracking
control problems for discrete-time systems. The continuous-time
counterpart problems have been solved in the literature, using a
Lyapunov method which has not been successfully used for discrete-time
systems.

On the other hand, the developed gradient algorithm framework
can be applied to adaptive state tracking control of
continuous-time systems, as illustrated next.

\bigskip
{\bf Direct adaptive control design}. The tracking error equation
(\ref{42.6m}), similar to (\ref{4erroreqm1}), is
\beq
\dot{e}(t) = A_m e(t) + B_m K_2^{*-1} \tilde{\Theta}^T(t) \omega(t),
\label{4erroreqm1c}
\eeq
and the continuous-time version of (\ref{4e(t)Wm(z)}) is
\beq
e(t) = W_m(s)[\left[ \begin{array}{c}
\rho_1^* \tilde{\theta}_1^T \omega\\
\rho_2^* \tilde{\theta}_2^T \omega\\
\vdots \\
\rho_M^* \tilde{\theta}_M^T \omega
\end{array}
\right]](t),
\label{4e(t)Wm(s)}
\eeq
for $W_m(s) = (sI - A_m)^{-1} B_m = [w_{ij}(s)]$ and $e(t) = [e_1(t),
  \ldots, e_n(t)]^T$. 

We can also introduce the estimation errors
\beq
\epsilon_i(t) = e_i(t) + \rho_{1}(t) \xi_{i1}(t)
+ \rho_{2}(t) \xi_{i2}(t) + \cdots + \rho_{M}(t) \xi_{iM}(t),
\label{4miepsi(t)mc}
\eeq
where $\rho_{j}(t)$, $j=1,2,\ldots, M$, are the estimates of
$\rho_{j}^*$, and
\beq
\xi_{ij}(t) = \theta_j^T(t) \zeta_{ij}(t) -
w_{ij}(s)[\theta_j^T \omega](t),\;\zeta_{ij}(t) =
w_{ij}(s)[\omega](t). 
\label{4mixii1mc}
\eeq
 
We then choose the adaptive laws
\bea
\label{4theta1t+1mc}
\dot{\theta}_i(t) \ts = \ts 
- \frac{\sign[\rho_{i}^*] \Gamma_i \sum_{k=1}^n \epsilon_k \zeta_{ki}}{m^2(t)}
\\*[0.05in]
\dot{\rho}_{i}(t) \ts = \ts 
- \frac{\gamma_i \sum_{k=1}^n \epsilon_k \xi_{ki}}{m^2(t)},
\label{4k22t+1mc}
\eea
where $\Gamma_i = \Gamma_i^T > 0$ and $\gamma_i > 0$,
 $i=1,2, \ldots, M$, and 
\beq
m(t) = \sqrt{1 + \sum_{i=1}^n \sum_{j=1}^M
(\zeta_{ij}^T(t) \zeta_{ij}(t) + \xi_{ij}^2(t))}. 
\eeq

\medskip
This adaptive scheme has the desired properties:

\begin{lem}
\label{4lemmadiec}
The adaptive laws (\ref{4theta1t+1mc})-(\ref{4k22t+1mc}) ensure:

(i) $\theta_i(t)$ and $\rho_i(t)$, $i=1,2, \ldots, M$, and
$\frac{\sum_{k=1}^n \epsilon_k^2(t)}{m^2(t)}$ are bounded; and 

(ii) $\frac{\sum_{k=1}^n \epsilon_k^2(t)}{m^2(t)} \in L^1$, 
$\frac{\epsilon_i(t)}{m(t)} \in L^2$, $\dot{\theta}_i(t) \in L^2$, 
$\dot{\rho}_i(t) \in L^{2}$.
\end{lem}

\bigskip
{\bf Indirect adaptive control design}. Based on the parametrized
plant equation (\ref{4param}):
\bea
\dot{x}(t) \ts = \ts A x(t) + B u(t) = A_m x(t) + B_m (\Theta_2^{*} u(t) -
\Theta_1^{*T} x(t)),
\eea
and the state estimator equation (\ref{4estm1}):
\beq 
\dot{\hat{x}}(t) = A_m \hat{x}(t) + B_m
(\Theta_2(t) u(t) - \Theta_1^{T}(t) x(t)) 
\eeq 
for the state estimation error $e_x(t) = \hat{x}(t) - x(t)$, we
obtained the error equation (\ref{4eeem1}):
\beq \dot{e}_x(t) = A_m e_x(t) + B_m ((\Theta_2(t) -
\Theta_2^*) u(t) - (\Theta_1(t) - \Theta_1^*)^T x(t)), 
\eeq
as the continuous-time version of (\ref{4eeem1d}), which can be expressed as
\beq
\dot{e}_x(t) = A_m e_x(t) + B_m \tilde{\Theta}^T(t) \omega(t),
\eeq
as similar to its discrete-time version (\ref{4erroreqmd}), and further expressed as
\beq
e_x(t) = W_m(s)[\left[ \begin{array}{c}
 \tilde{\theta}_1^T \omega\\
 \tilde{\theta}_2^T \omega\\
\vdots \\
 \tilde{\theta}_M^T \omega
\end{array}
\right]](t),
\eeq
as similar to its discrete-time version (\ref{4erroreqmd1}).

For $e_x(t) = [e_{x1}(t), e_{x2}(t), \ldots, e_{xn}(t)]^T$ and $W_m(s) =
[w_{ij}(s)]$, $i=1,2,\ldots, n$, $j = 1, 2, \ldots, M$, we can
similarly introduce the estimation errors
\beq
\epsilon_i(t) = e_{xi}(t) + \xi_{i1}(t)
+ \xi_{i2}(t) + \cdots + \xi_{iM}(t),
\label{4miepsi(t)mdc}
\eeq
where
\beq
\xi_{ij}(t) = \theta_j^T(t) \zeta_{ij}(t) -
w_{ij}(s)[\theta_j^T \omega](t),\;\zeta_{ij}(t) =
w_{ij}(s)[\omega](t). 
\label{4mixii1mdc}
\eeq
 
For $i=1,2,\ldots, M$, we choose the adaptive laws
\beq
\dot{\theta}_i(t) = 
- \frac{\Gamma_i \sum_{k=1}^n \epsilon_k \zeta_{ki}}{m^2(t)}
\label{4theta1t+1mdc}
\eeq
where $\Gamma_i = \Gamma_i^T > 0$, and
\beq
m(t) = \sqrt{1 + \sum_{i=1}^n \sum_{j=1}^M
(\zeta_{ij}^T(t) \zeta_{ij}(t) + \xi_{ij}^2(t))}. 
\eeq

\medskip
This adaptive scheme has the similar properties to that in
Lemma \ref{4lemmadie}:

\medskip
\medskip
(i) $\theta_i(t)$, $i=1,2, \ldots, M$, and
$\frac{\sum_{k=1}^n \epsilon_k^2(t)}{m^2(t)}$ are bounded; and 

\medskip
(ii) $\frac{\sum_{k=1}^n \epsilon_k^2(t)}{m^2(t)} \in L^1$, 
$\frac{\epsilon_i(t)}{m(t)} \in L^2$, and $\dot{\theta}_i(t) \in L^{2}$.

\medskip
Parameter projection can also be used to ensure that the parameter
matrix $\Theta_2(t)$ in $\Theta(t) = [\Theta_1^T(t), \Theta_2(t)]^T$
is nonsingular, for calculating the parameters 
\beq
K_1^T(t) = \Theta_2^{-1}(t)
\Theta_1^T(t),\;K_2(t) = \Theta_2^{-1}(t),
\eeq
to implement the adaptive control law
\beq
u(t) = K_1^T(t) x(t) + K_2(t) r(t),
\label{4aclmdc}
\eeq

{\bf Discussion}. Based on the desired adaptive parameter estimation
properties (see Lemma \ref{4lemmadiec}), similar to the procedure for
the proof of Theorem \ref{4thm}, a continuous-time version of the
operator-based theory can be derived to establish the closed-loop
signal boundedness and asymptotic tracking of $x_m(t)$ by $x(t)$ for
the new gradient algorithm based continuous-time adaptive state
tracking schemes developed in Section \ref{4Application to
  Continuous-Time Systems}.

Such gradient algorithm based adaptive state tracking control schemes
are new additions to the Lyapunov algorithm based continuous-time
adaptive state tracking schemes presented in Section \ref{4Designs for
  Continuous-Time SystemsMIMO} (their single-input versions developed
in Section \ref{4Discrete-Time Adaptive Control Designs} are new 
additions to that presented in Section \ref{4Continuous-Time
  Designs}), to expand the solutions to the adaptive state tracking
control problems.

\section{Concluding Remarks}
In this paper, we have studied a new gradient algorithm based framework
for adaptive state tracking control of a continuous-time system:
$\dot{x}(t) = A x(t) + B u(t)$ or a discrete-time system: $x(t+1) = A
x(t) + B u(t)$, for the state vector $x(t)$ to asymptotically track
the state vector $x_m(t)$ of a chosen and stable reference model
system. The gradient algorithm based framework has been used to develop new
direct adaptive control and indirect adaptive control schemes, either
to solve the open discrete-time state tracking control problem, or to
provide new solutions to the continuous-time adaptive state
tracking control problem which was solved in the literature with
a Lyapunov method based framework (but its applicability to
discrete-time systems has not been verified).

\end{document}